\documentclass[preprint,aps,prd,showpacs,nofootinbib,tightenlines]{revtex4}
\usepackage[colorlinks,urlcolor=blue,linkcolor=blue,citecolor=blue]{hyperref} %
\usepackage{amsmath}
\usepackage{amssymb}
\usepackage{epsfig}
\usepackage{graphicx}
\usepackage{float}
\usepackage{color} 
\usepackage{booktabs}
\usepackage{ulem}  
\usepackage{verbatim} 
\def \KorKbar{{\raisebox{8.5pt}[0pt][0pt]{\fontsize{2.0pt}{\baselineskip}\selectfont $($}}$\overline K${\raisebox{8.5pt}[0pt][0pt]{\fontsize{2.0pt}{\baselineskip}\selectfont $)$}}}  
\begin{document}
\title{\Large{Contributions of $K^*_0(1430)$ and $K^*_0(1950)$ in the three-body decays $B\to K\pi h$}\vspace{0.4cm}} 

\author{Wen-Fei Wang$^{1,2}$}\email{wfwang@sxu.edu.cn}
\author{Jian Chai$^{1,2}$}
\author{Ai-Jun Ma$^{3}$}

\affiliation{$^1$Institute of Theoretical Physics, Shanxi University, Taiyuan, Shanxi 030006, China}
\affiliation{$^2$State Key Laboratory of Quantum Optics and Quantum Optics Devices, Shanxi University, 
                        Taiyuan, Shanxi 030006, China}
\affiliation{$^{3}$Department of Mathematics and Physics,  Nanjing Institute of Technology, Nanjing, Jiangsu 211167, China}  
                        
\date{\today}
\begin{abstract}
We study the contributions of the resonant states $K_0^{*}(1430)$ and $K_0^{*}(1950)$ in the three-body decays 
$B\to K\pi h$ (with $h=\pi, K$) in the perturbative QCD approach. The crucial nonperturbative input 
$F_{K\pi}(s)$ in the distribution amplitudes of the $S$-wave $K\pi$ system is derived from the matrix element of 
vacuum to $K\pi$ pair. The $CP$ averaged  branching fraction of the quasi-two-body decay process 
$B\to K^*_0(1950)h\to K\pi h$ is about one order smaller than that of the corresponding decay $B\to K^*_0(1430)h\to K\pi h$. 
In view of the important contribution from the $S$-wave $K\pi$ system for the $B\to K\pi h$ decays, 
it is not appropriate to neglect the $K_0^{*}(1950)$ in the theoretical or experimental studies for the relevant three-body 
$B$ meson decays. The predictions in this work for the relevant decays are consistent with the existing experimental data.
\end{abstract}
\maketitle
\section{Introduction}

The charmless three-body hadronic $B$ meson decay processes provide us a field to appraise different dynamical models of 
strong interaction, to investigate hadronic final-state interactions and analyze hadron spectroscopy, to determine the fundamental 
quark mixing parameters and understand $CP$ asymmetries. In order to extract the significative information from experimental 
results and present the effective and accurate predictions for the three-body $B$ decays, some methods have 
been adopted in abundant works, such as the $U$-spin, isospin and flavor $SU(3)$ symmetries in~\cite{plb564-90,prd72-075013,
prd72-094031,prd84-056002,plb727-136,plb726-337,prd89-074043,plb728-579,ijmpa29-1450011,prd91-014029}, 
the QCD factorization (QCDF) in~\cite{plb622-207,prd74-114009,prd79-094005,prd81-094033,appb42-2013,prd72-094003,
prd76-094006,prd88-114014,prd89-074025,prd89-094007,prd94-094015,epjc75-536,prd87-076007,epjc78-845,npb899-247} 
and the perturbative QCD (PQCD) approach in~\cite{plb561-258,prd70-054006,prd89-074031}. The three-body decays 
$B\to K\pi h$, with $h$ is the pion or kaon, have been studied by Belle~\cite{prd71-092003,prl96-251803,
prd75-012006,prd79-072004,prd96-031101,prd99-031102,prd100-011101}, BaBar~\cite{prd72-072003,prl99-221801,prd78-012004,
prd78-052005,prd80-112001,prd82-031101,prd83-112010,prd96-072001} and LHCb~\cite{jhep1310-143,
prl111-101801,prl112-011801,prd90-112004,plb765-307,jhep1711-027,prl120-261801,jhep1906-114} Collaborations in recent years.
These decays especially the $B\to K\pi\pi$ were found to be a clean source for the extraction of the Cabibbo-Kobayashi-Maskawa 
(CKM)~\cite{CKM-C,CKM-KM} angle $\gamma$~\cite{prd74-051301,prd75-014002,prd76-073011,plb645-201,prd84-034041,
prd85-016010,plb728-206,jhep1504-154}. The relevant processes also provide new possibilities for the measurements of the 
$CP$ violation in the $B$ decays~\cite{prl96-251803,prl111-101801,prl112-011801,prd90-112004}. 

The total decay amplitude for the $B$ meson decays into three light mesons $K,~\pi$ and $h$ as the final state can be 
described as the coherent sum of the nonresonant and resonant contributions in the isobar formalism~\cite{pr135-B551,pr166-1731,
prd11-3165}. The nonresonant contributions are spread all over the phase space and play an important role in the corresponding 
decay processes~\cite{prd60-054029,prd66-054015,plb665-30}. The resonant contributions from low energy scalar, vector and 
tensor resonances are known experimentally, in most cases, to be the dominated proportion of the related decays and could be 
studied in the quasi-two-body framework~\cite{plb763-29,1605-03889,prd96-113003} when the rescattering 
effects~\cite{1512-09284} and three-body effects~\cite{npps199-341,prd84-094001} are neglected.  For the three-body decays 
$B\to K\pi h$, one has the resonant contributions from the $K\pi,~\pi h$ and $Kh$ pairs which are originated from 
different intermediate states and as well containing the two-body final state interactions. And the $J^P=0^+$ component of the 
$K\pi$ spectrum, denoted as $(K\pi)^{*}_0$, is always found very important for the relevant physical observables.  

The kaon-pion scattering has been extensively studied in 
Refs~\cite{epjc77-91,prd93-074025,prd123-042002,npb932-29,prd91-054008,prd86-054508,prd92-113002} in recent years.
While the primary source of the information on $I=1/2~S$-wave $K\pi$ system comes from the LASS experiment for the reaction 
$K^-p \to K^-\pi^+n$~\cite{npb296-493}. The $K\pi$ $S$-wave amplitude has also been studied in detail in the decays 
$D^+\to K^-\pi^+\pi^+$ by E791~\cite{prd73-032004}, FOCUS~\cite{plb653-1,plb681-14} and CLEO~\cite{prd78-052001}, 
$\eta_c\to K \bar K\pi$ by BaBar~\cite{prd93-012005} and $\tau^-\to K_S\pi^-\nu_\tau$ by Belle~\cite{plb654-65} with the 
methods of Breit-Wigner functions~\cite{BW-model}, K-matrix formalism~\cite{pr70-15,npa189-417,ap4-404} or 
model-independent partial-wave analysis. To describe the slowly increasing phase as a function of the $K\pi$ invariant mass, the scalar 
$K\pi$ scattering amplitude was written as the relativistic Breit-Wigner term~\cite{BW-model} for the resonance $K^*_0(1430)$ in the 
LASS parametrization together with an effective range nonresonant component in~\cite{npb296-493}, and the effective range term has 
been applied a cutoff to the slowly varying part close to the charm hadron mass at about $1.8$ GeV for the three-body $B$ decays in 
the experimental studies~\cite{prd78-012004,prd80-112001,prd83-112010,prd96-072001}.  At about $1.95$ GeV one will find the 
presence of the resonance $K^*_0(1950)$ in~\cite{npb296-493} and also in the $\eta_c$ decays in~\cite{prd93-012005,prd89-112004}.
This state was assigned as a radial excitation of the $0^+$ member of the $L=1$ triplet in the LASS analysis~\cite{npb296-493}.  
The lowest-lying broad component of the $S$-wave $K\pi$ system is the $K^*_0(700)$~\cite{PDG-2018}, 
also named as $\kappa$ or $K^*_0(800)$ in 
literature~\cite{prl89-121801,plb632-471,plb633-681,prd73-032004,epjc48-553,prd78-052001,plb681-14,prd81-014002,plb698-183},  
which has commonly been placed together with the resonant states $\sigma$, $f_0(980)$ and $a_0(980)$ into an $SU(3)$ flavor nonet, 
and they have been suspected to be exotics~\cite{prd15-267,prl48-659,prd41-2236,prl93-212002,pr389-61,pr409-1,prd74-014028}.

In this work, we will focus on the contributions of the resonant state $K^*_0(1430)$  in the $B\to K\pi h$ decay processes in the 
PQCD approach based on the $k_{\rm T}$ factorization theorem~\cite{plb504-6,prd63-054008,prd63-074009,ppnp51-85}. The 
contributions of the resonant state $K^*_0(1950)$ in the hadronic three-body $B$ meson decays involving $S$-wave $K\pi$ pair 
have been ignored in the relevant theoretical studies and be noticed only by LHCb Collaboration very recently in the 
works~\cite{prd90-072003,epjc78-1019}. 
We will systematically estimate, for the first time, the contributions from the state $K^*_0(1950)$ for the $B\to K\pi h$ decays in this 
work. As for the resonance $K^*_0(700)$, we shall leave to the future studies in view of its ambiguous internal structure and the accompanying complicated results for the three-body $B$ decays~\cite{prd73-014017}, in addition, the corresponding 
contributions have been covered up by the effective range part of LASS line shape for the experimental results~\cite{prd78-012004,
prd78-052005,prd80-112001,prd83-112010,prd96-072001}. 

For the quasi-two-body decays $B\to K^*_0(1430,1950)h \to K\pi h$, the subprocesses of the $B\to K\pi h$ decays, the intermediate 
state $K^*_0$, as demonstrated in the Fig.~\ref{fig-feyndiag}, is generated in the hadronization of quark-antiquark pair including 
one $s$ or $\bar s$-quark. The process $K^*_0\to K \pi$, which can not be calculated in the PQCD approach, is always shrunken 
as the decay constants in the twist-2 and twist-3 light-cone distribution amplitudes of the scalar 
mesons~\cite{prd73-014017,prd75-056001,epja49-78} in the studies of the two-body $B$ meson decays involving the scalar 
mesons $K^*_0(700)$ and $K^*_0(1430)$, see Ref.~\cite{prd95-016011} and the references therein for examples.
While in the quasi-two-body framework based on PQCD, one can easily introduce the nonperturbative subprocess 
$K^*_0\to K \pi$ into a time-like form factor in the distribution amplitudes of the $K\pi$ pair.
The quasi-two-body framework based on PQCD has been discussed in detail in~\cite{plb763-29} and has been adopted 
in some studies on the quasi-two-body $B$ meson decay processes recently~\cite{prd100-014017,epjc79-37,epjc79-539,
plb791-342,plb788-468,prd96-093011,prd96-036014,epjc77-199,prd95-056008,npb923-54}.

This work is organized as follows. In Sec.~II, we give a brief introduction for the theoretical framework.
In Sec.~III, we show the numerical results and give some discussions. Conclusions are presented in Sec.~IV.
The factorization formulas and functions for the related quasi-two-body decay amplitudes are collected in the Appendix.

\section{Framework}

\begin{figure}[tbp]
\centerline{\epsfxsize=14cm \epsffile{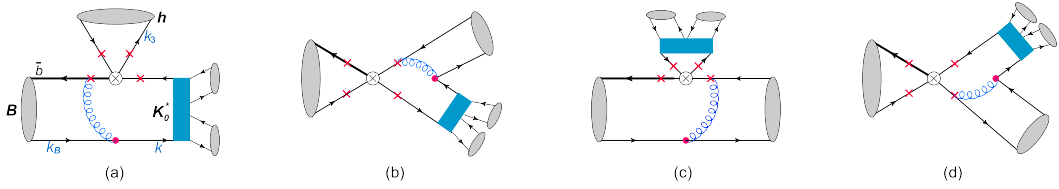}}
\caption{Typical Feynman diagrams for the decay processes $B\to K_0^* h\to K\pi h$, $h=$($\pi, K$). The symbol $\otimes$ 
               is the weak vertex, $\times$ denotes possible attachments of hard gluons and the rectangle represents the scalar 
               resonances $K_0^*$.}
\label{fig-feyndiag}
\end{figure}

In the rest frame of $B$ meson, we define its momentum $p_B$ and light spectator quark momentum $k_B$ as
\begin{eqnarray}
p_B=\frac{m_B}{\sqrt2}(1,1,0_{\rm T}),\quad  k_B=\left(\frac{m_B}{\sqrt2}x_B, 0, k_{B{\rm T}}\right),
\end{eqnarray}
in the light-cone coordinates, where $x_B$ is the momentum fraction and $m_B$ is the mass. For the resonant states 
$K^*_0$ and the $K\pi$ pair generated from it by the strong interaction as revealed in the Fig.~\ref{fig-feyndiag}, we define 
their momentum $p=\frac{m_B}{\sqrt 2}(\zeta, 1, 0)$. Its easy to validate $\zeta=s/m_B^2$, where the invariant mass square 
$s=p^2=m^2_{K\pi}$ for the $K\pi$ pair.  The light spectator quark comes from $B$ meson 
and goes into intermediate state in the hadronization of $K^*_0$  as shown in Fig.~\ref{fig-feyndiag}~(a) has the momentum 
$k=(0, \frac{m_B}{\sqrt 2}z, k_{\rm T})$. 
For the bachelor final state $h$ and its spectator quark, their momenta $p_3$ and $k_3$ have the definitions as
\begin{eqnarray}
p_3=\frac{m_B}{\sqrt2}(1-\zeta, 0, 0_{\rm T}),\quad
k_3=\left(\frac{m_B}{\sqrt2}(1-\zeta)x_3, 0, k_{3{\rm T}}\right).
\end{eqnarray}
Where $x_3$ and $z$, which run from 0 to 1, are the corresponding momentum fractions. 

The matrix element from the vacuum to the $K^+\pi^-$ final state is given by~\cite{zpc29-637}
\begin{eqnarray}
\langle K^+(p_1)\pi^-(p_2)|\bar d\gamma_\mu (1-\gamma_5) s|0\rangle
=\left[(p_1-p_2)_\mu- \frac{\Delta_{K\pi}}{p^2}p_\mu \right] F^{K\pi}_+(s) + \frac{\Delta_{K\pi}}{p^2}p_\mu F^{K\pi}_0(s),\;\;
\end{eqnarray}
with the $p_1(p_2)$ is the momentum for kaon(pion) in the $K\pi$ system, $\Delta_{K\pi}=(m^2_K-m^2_\pi)$ and 
$m_K(m_\pi)$ is the mass of $K(\pi)$ meson.
The $F^{K\pi}_+(s)$ is the vector form factor which has been discussed in detail in the 
Refs.~\cite{plb654-65,epjc11-599,plb640-176,plb664-78,epjc53-401,epjc59-821,jhep1009-031,jhep1409-042,jhep1912-083}.
The the scalar form factor $F^{K\pi}_0(s)$ is defined as~\cite{npb622-279,prd80-054007,plb730-336}
\begin{eqnarray} 
\langle K\pi | \bar{q} s| 0 \rangle =C_X \frac{\Delta_{K\pi}}{m_s-m_{q}} F_0^{K\pi}(s)= B_0 C_X F_0^{K\pi}(s)\;,
\end{eqnarray}
where $q$ is the light quark $u$ or $d$, the isospin factor $C_X=1$ for $X=\{K^+\pi^-, K^0\pi^+\}$ and $C_X=1/\sqrt{2}$ for 
$X=\{K^+\pi^0, K^0\pi^0\}$. The constant $B_0$ equals to $\Delta_{K\pi}/(m_s-m_{q})$. The form factor $F^{K\pi}_0(s)$ above is 
suppose to be one when $s$ goes to zero. When the $K^+\pi^-$ pair originated from the resonant state $K^*_0(1430)^0$, we 
have~\cite{prd80-054007}
\begin{eqnarray}
\langle K^+\pi^- | \bar d s| 0 \rangle \approx  \langle K^+\pi^- |  K_0^{*0} \rangle \frac{1}{\mathcal{D}_{K^*_0}} 
 \langle  K_0^{*0} |\bar d s |0 \rangle = \Pi_{K_0^*K\pi} \langle  K_0^{*0} |\bar d s |0 \rangle       \;,
\end{eqnarray}
and
\begin{eqnarray}
\Pi_{K_0^*K\pi}=\frac{g_{K_0^*K\pi}}{\mathcal{D}_{K^*_0}} 
\approx \frac{B_0 }{\bar f_{K_0^*} m_{K_0^*}} F_0^{K\pi}(s)\;,  
\label{eq-vert-SKpi}
\end{eqnarray}
with $\bar f_{K_0^*}= \frac{m_{K_0^*}}{m_s(\mu)-m_d(\mu)}\cdot f_{K_0^*}$, the decay constants defined by 
$\langle  K_0^{*0} |\bar d s |0 \rangle=m_{K_0^*}\bar f_{K_0^*}$ and $\langle  K_0^{*0}(p) |\bar d\gamma_\mu s|0 \rangle=
f_{K_0^*}p_\mu$~\cite{prd73-014017}, and the mass $m_{K_0^*}$ could be replaced by the invariant mass $\sqrt s$ for the off-shell 
$K_0^{*}$.  One can find different values of $f_{K_0^*}$ for $K_0^*(1430)$ in~\cite{prd78-114016}, we employ 
$f_{K_0^*(1430)}m^2_{K_0^*(1430)}=0.0842\pm0.0045$ GeV$^3$~\cite{plb462-14} and $f_{K_0^*(1950)}m^2_{K_0^*(1950)}=0.0414$ 
GeV$^3$~\cite{prd63-074017} in this work.
The Breit-Wigner formula for the denominator ${\mathcal{D}_{K^*_0}}=m^2_{K^*_0}-s-i m_{K^*_0}\Gamma(s)$, with the 
mass-dependent decay width $\Gamma(s)=\Gamma_0\frac{q}{q_0}\frac{m_{K^*_0}}{\sqrt s}$ and $\Gamma_0$ is the full width 
for resonant state $K_0^{*}$. In the rest frame of the resonance $K_0^{*}$, its daughter kaon or pion has the magnitude of the 
momentum as
\begin{eqnarray}
q=\frac{1}{2}\sqrt{\left[s-(m_K+m_\pi)^2\right]\left[s-(m_K-m_\pi)^2\right]/s}\;.
\end{eqnarray}
The $q_0$ in $\Gamma(s)$ is the value for $q$ at $s=m^2_{K_0^{*}}$. 
The coupling constant $g_{K_0^*K\pi}=\langle K^+\pi^- | K_0^{*0} \rangle$, one has~\cite{prd88-114014}  
\begin{eqnarray}
g_{K_0^*K\pi}=\sqrt{\frac{8\pi m^2_{K^*_0} \Gamma_{{K^*_0}\to K\pi}}{q_0}}\;,
\label{eq-g-SKpi}
\end{eqnarray}
where the $\Gamma_{{K^*_0}\to K\pi}$ is the partial width for ${K^*_0}\to K\pi$.

The $S$-wave $K\pi$ system distribution amplitudes are collected into~\cite{plb730-336,prd73-014017,prd87-114001,prd91-094024}
\begin{eqnarray}
\Phi_{K\pi}(z,s)=\frac{1}{\sqrt{2N_c}}\left[{p\hspace{-1.6truemm}/}\phi(z,s)
+\sqrt s\phi^{s}(z,s)+\sqrt s({v \hspace{-1.7truemm}/ }{n \hspace{-1.8truemm}/}-1)
\phi^{t}(z,s) \right],
\label{def-wavefun-Kpi}
\end{eqnarray}
with the $v=(0,1,{0}_{\rm T})$ and $n=(1,0,{0}_{\rm T})$ being the dimensionless vectors.~The twist-2 light-cone distribution amplitude 
has the form~\cite{plb730-336,prd73-014017,prd87-114001}
\begin{eqnarray}
\phi(z,s)=\frac{F_{K\pi}(s) }{2\sqrt{2N_c}}\left\{ 6z(1-z)\left[a_{0}(\mu)+
\sum^{\infty}_{m=1} a_{m}(\mu)C^{3/2}_m(2z-1) \right]\right\}\;,
\label{def-wavefun-twist2}
\end{eqnarray}
with $C^{3/2}_m$ the Gegenbauer polynomials, $a_0=(m_s(\mu)-m_q(\mu))/\sqrt s$ for ($K^{*-}_0, \bar K^{*0}_0$) and $a_0=(m_q(\mu)-m_s(\mu))/\sqrt s$ for  ($K^{*+}_0, K^{*0}_0$) according to Ref.~\cite{prd87-114001}. 
The $a_m$ are scale-dependent Gegenbauer moments, with $a_1=-0.57\pm0.13$ and $a_3=-0.42\pm0.22$ at the scale 
$\mu=1$ GeV for the resonance $K^*_0(1430)$, and the contributions from the even terms could be neglected~\cite{prd73-014017}.  
There is no available Gegenbauer moments for the state $K^*_0(1950)$, we employ the scale-dependent $a_1$ and $a_3$ of 
$K^*_0(1430)$ for the entire $S$-wave $K\pi$ system in the numerical calculation.
For the twist-3 light-cone distribution amplitudes in this work, we take the asymptotic forms as
\begin{eqnarray}
\phi^{s}(z,s)=\frac{F_{K\pi}(s) }{2\sqrt{2N_c}}\;,\quad\quad
\phi^{t}(z,s)=\frac{F_{K\pi}(s) }{2\sqrt{2N_c}}(1-2z)\;.
\label{def-wavefun-twist3}
\end{eqnarray}
The factor $F_{K\pi}(s)$ is related to scalar form factor $F^{K\pi}_0(s)$ by $F_{K\pi}(s)= \frac{B_0 }{m_{K_0^*}} F_0^{K\pi}(s)$.

The distribution amplitudes for $B$ meson and the bachelor final state $h$ in this work are the same as those widely employed in 
the studies of the hadronic $B$ meson decays in the PQCD approach, one can find their expressions and parameters in the 
Appendix. 

\section{Results and discussions}
 
In the numerical calculation, we adopt the decay constants $f_B=0.189$ GeV, $f_{B_s}=0.231$ GeV~\cite{prd98-074512}, 
the mean lifetimes $\tau_{B^0}=(1.520\pm0.004)\times 10^{-12}$~s, $\tau_{B^+}=(1.638\pm0.004)\times 10^{-12}$~s and 
 $\tau_{B^0_s}=(1.509\pm0.004)\times 10^{-12}$~s~\cite{PDG-2018} for the $B^0, B^+$ and $B^0_s$ mesons, respectively. 
The masses and the decay constants for the relevant particles in the numerical calculation in this work, the full widths for 
$K^*_0(1430)$ and $K^*_0(1950)$, and the Wolfenstein parameters of the CKM matrix are presented in Table~\ref{tab1}.

\begin{table}[H]  
\begin{center}
\caption{Masses, decay constants, full widths of $K^*_0(1430)$ and $K^*_0(1950)$ (in units of GeV) and Wolfenstein 
parameters~\cite{PDG-2018}.}
\label{tab1}
\begin{tabular}{l}\hline\hline 
\;$m_{B^{0}}=5.280 \quad\;\; m_{B^{\pm}}=5.279 \quad\; m_{B^0_s}=5.367 \quad\; m_{\pi^\pm}=0.140   
\quad\; m_{\pi^0}=0.135 \;$ \\ \vspace{0.1cm} 
\;$m_{K^\pm}=0.494 \quad\;  m_{K^0}=0.498 \quad\;\; f_K=0.156\quad\;\;\; f_\pi=0.130\; $ \\ 
\;$m_{K^*_0(1430)}=1.425\pm0.050  \hspace{1.95cm} \Gamma_{K^*_0(1430)}=0.270\pm0.080\;$  \vspace{-0.05cm} \\ 
\;$m_{K^*_0(1950)}=1.945\pm0.010\pm0.020 \quad\;\;\Gamma_{K^*_0(1950)}=0.201\pm0.034\pm0.079\;$  \vspace{0.12cm}\\
\;$\lambda=0.22453\pm 0.00044  \;\;\;  A=0.836\pm0.015  \;\;\;  \bar{\rho} = 0.122^{+0.018}_{-0.017}  
\;\;\; \bar{\eta}= 0.355^{+0.012}_{-0.011} $\\ 
\hline\hline  
\end{tabular}
\end{center}
\end{table}

Utilizing the differential branching fraction Eq.~(\ref{eqn-diff-bra}) and the decay amplitudes collected in 
Appendix~\ref{sec-appx-a}, we obtain the 
$CP$ averaged branching fractions (${\mathcal B}$) and the direct $CP$ asymmetries (${\mathcal A}_{CP}$) in Table~\ref{Res-1430} 
and Table~\ref{Res-1950} for the concerned quasi-two-body decay processes involving the resonances $K_0^{*}(1430)$ and 
$K_0^{*}(1950)$ as the intermediate states, respectively. The results for those quasi-two-body decays with one 
daughter of the $K_0^{*}$ is the neutral pion are omitted. One will get a half value of the ${\mathcal B}$ and the same value 
of the ${\mathcal A}_{CP}$ of the corresponding result in Tables~\ref{Res-1430}, \ref{Res-1950} for a decay with the subprocesses 
$K_0^{*}\to K \pi^0$ considering the isospin relation. For example, we have
\begin{eqnarray}
{\mathcal B}(B^+\to K_0^{*}(1430)^+\pi^0\to K^+\pi^0 \pi^0)=\frac12{\mathcal B}(B^+\to K_0^{*}(1430)^+\pi^0\to K^0\pi^+ \pi^0), 
\end{eqnarray}
while these two processes have the same direct $CP$ asymmetry.

\begin{table}[thb]   
\begin{center}
\caption{PQCD predictions of the $CP$ averaged branching fractions and the direct $CP$
         asymmetries for the quasi-two-body $B\to K_0^*(1430)h \to K\pi h$ decays.}
\label{Res-1430}   
\begin{tabular}{l c l} \hline\hline
   \quad Decay modes       &    ~    &  \quad\; Quasi-two-body results               \\
\hline  %
  $B^+\to  K_0^{*}(1430)^0 \pi^+\to K^+\pi^- \pi^+$    &${\mathcal B}(10^{-5})$\;
      & $2.27\pm0.59(\omega_B)\pm0.17(a_{3+1})\pm0.34(m^\pi_0{+}a^\pi_2)$    \\
              &  ${\mathcal A}_{CP}(\%)$\;
      & $-1.3\pm0.2(\omega_B)\pm0.4(a_{3+1})\pm0.2(m^\pi_0{+}a^\pi_2)$   \\
  $B^+\to K_0^{*}(1430)^+ \pi^0\to K^0\pi^+ \pi^0$    &${\mathcal B}(10^{-6})$\;
      & $7.86\pm2.16(\omega_B)\pm0.55(a_{3+1})\pm1.36(m^\pi_0{+}a^\pi_2)$    \\
               &  ${\mathcal A}_{CP}(\%)$\;
      & $1.5\pm0.4(\omega_B)\pm0.8(a_{3+1})\pm0.4(m^\pi_0{+}a^\pi_2)$   \\
  $B^+\to K_0^{*}(1430)^+ \bar K^0\to K^0\pi^+ \bar K^0$    &${\mathcal B}(10^{-7})$\;
   & $2.33\pm0.04(\omega_B)\pm1.29(a_{3+1})\pm0.34(m^K_0{+}a^K_2)$ \\
              &  ${\mathcal A}_{CP}(\%)$\;
          & $-18.4\pm5.8(\omega_B)\pm2.7(a_{3+1})\pm5.4(m^K_0{+}a^K_2)$ \\
  $B^+\to \bar K_0^{*}(1430)^0 K^+\to K^-\pi^+ K^+$     &${\mathcal B}(10^{-6})$\;   
      & $2.86\pm0.54(\omega_B)\pm0.51(a_{3+1})\pm0.42(m^K_0{+}a^K_2)$    \\
               &  ${\mathcal A}_{CP}(\%)$\;
      & $17.9\pm0.4(\omega_B)\pm8.0(a_{3+1})\pm0.9(m^K_0{+}a^K_2)$\   \\
\hline %
  $B^0\to K_0^{*}(1430)^+ \pi^-\to K^0\pi^+ \pi^-$ &${\mathcal B}(10^{-5})$
      & $2.07\pm0.54(\omega_B)\pm0.14(a_{3+1})\pm0.30(m^\pi_0{+}a^\pi_2)$    \\
              &  ${\mathcal A}_{CP}(\%)$\;
      & $0.3\pm0.5(\omega_B)\pm0.8(a_{3+1})\pm0.1(m^\pi_0{+}a^\pi_2)$    \\
  $B^0\to  K_0^{*}(1430)^0 \pi^0\to K^+\pi^- \pi^0$ &${\mathcal B}(10^{-5})$\;
      & $1.39\pm0.35(\omega_B)\pm0.11(a_{3+1})\pm0.18(m^\pi_0{+}a^\pi_2)$    \\
              &  ${\mathcal A}_{CP}(\%)$\;
      & $-1.8\pm0.4(\omega_B)\pm0.2(a_{3+1})\pm0.1(m^\pi_0{+}a^\pi_2)$   \\
  $B^0\to K_0^{*}(1430)^+  K^-\to K^0\pi^+ K^-$\    &${\mathcal B}(10^{-8})$\;
      & $5.77\pm2.38(\omega_B)\pm2.92(a_{3+1})\pm0.62(m^K_0{+}a^K_2)$   \\
              &  ${\mathcal A}_{CP}(\%)$\;
      & $4.9\pm6.4(\omega_B)\pm3.7(a_{3+1})\pm3.6(m^K_0{+}a^K_2)$    \\ 
  $B^0\to K_0^{*}(1430)^- K^+\to \bar K^0\pi^- K^+$     &${\mathcal B}(10^{-7})$\;
      & $3.84\pm1.48(\omega_B)\pm1.95(a_{3+1})\pm0.09(m^K_0{+}a^K_2)$   \\
               &  ${\mathcal A}_{CP}(\%)$\;
      & $-5.0\pm2.6(\omega_B)\pm6.7(a_{3+1})\pm3.0(m^K_0{+}a^K_2)$ \\
  $B^0\to  K_0^{*}(1430)^0 \bar K^0\to K^+\pi^- \bar K^0$ &${\mathcal B}(10^{-7})$\;
      & $3.04\pm0.15(\omega_B)\pm2.04(a_{3+1})\pm0.36(m^K_0{+}a^K_2)$   \\
              &  ${\mathcal A}_{CP}(\%)$\;
      & \quad - \\
  $B^0\to \bar K_0^{*}(1430)^0 K^0\to K^-\pi^+ K^0$ &${\mathcal B}(10^{-6})$\;
      & $2.89\pm0.53(\omega_B)\pm0.65(a_{3+1})\pm0.41(m^K_0{+}a^K_2)$   \\
              &  ${\mathcal A}_{CP}(\%)$\;
      & \quad - \\
\hline %
  $B_s^0\to  K_0^{*}(1430)^-  \pi^+\to \bar K^0\pi^- \pi^+$ &${\mathcal B}(10^{-5})$\;
      & $3.77\pm0.78(\omega_B)\pm0.51(a_{3+1})\pm0.01(m^\pi_0{+}a^\pi_2)$    \\
              &  ${\mathcal A}_{CP}(\%)$\;
      & $15.5\pm1.6(\omega_B)\pm3.2(a_{3+1})\pm1.0(m^\pi_0{+}a^\pi_2)$     \\
  $B_s^0\to \bar K_0^{*}(1430)^0 \pi^0\to K^-\pi^+ \pi^0$ &${\mathcal B}(10^{-7})$\;
      & $5.03\pm0.38(\omega_B)\pm1.52(a_{3+1})\pm0.80(m^\pi_0{+}a^\pi_2)$    \\
              &  ${\mathcal A}_{CP}(\%)$\;
      & $59.2\pm3.2(\omega_B)\pm7.1(a_{3+1})\pm2.5(m^\pi_0{+}a^\pi_2)$    \\
  $B_s^0\to  K_0^{*}(1430)^+  K^-\to K^0\pi^+  K^-$ &${\mathcal B}(10^{-5})$\;
      & $1.44\pm0.20(\omega_B)\pm0.26(a_{3+1})\pm0.25(m^K_0{+}a^K_2)$    \\
              &  ${\mathcal A}_{CP}(\%)$\;
      & $0.4\pm0.3(\omega_B)\pm1.8(a_{3+1})\pm1.2(m^K_0{+}a^K_2)$   \\
  $B_s^0\to K_0^{*}(1430)^- K^+\to \bar K^0\pi^- K^+$ &${\mathcal B}(10^{-5})$\;
      & $1.74\pm0.16(\omega_B)\pm0.84(a_{3+1})\pm0.24(m^K_0{+}a^K_2)$   \\
              &  ${\mathcal A}_{CP}(\%)$\;
      & $-51.1\pm1.1(\omega_B)\pm6.7(a_{3+1})\pm5.5(m^K_0{+}a^K_2)$    \\
  $B_s^0\to  K_0^{*}(1430)^0 \bar K^0\to K^+\pi^- \bar K^0$ &${\mathcal B}(10^{-5})$\;
      & $1.47\pm0.22(\omega_B)\pm0.24(a_{3+1})\pm0.25(m^K_0{+}a^K_2)$   \\
              &  ${\mathcal A}_{CP}(\%)$\;
      & \quad -    \\
  $B_s^0\to \bar K_0^{*}(1430)^0 K^0\to K^-\pi^+ K^0$ &${\mathcal B}(10^{-5})$\;
      & $1.19\pm0.06(\omega_B)\pm0.71(a_{3+1})\pm0.17(m^K_0{+}a^K_2)$   \\
              &  ${\mathcal A}_{CP}(\%)$\;
      & \quad -   \\
\hline\hline
\end{tabular}
\end{center}
\end{table}

For the PQCD predictions in Tables~\ref{Res-1430}, \ref{Res-1950}, the shape parameters $\omega_B=0.40\pm0.04$ or 
$\omega_{B_s}=0.50 \pm 0.05$ in Eq.~(\ref{phib}) for the $B^{+,0}$ or $B^0_s$ contribute the first error. The second error for 
each PQCD result comes from the Gegenbauer moments $a_1$ and $a_3$ in the Eq.~(\ref{def-wavefun-twist2}). 
The third one is induced by the chiral masses $m^h_0$ and the Gegenbauer moment $a^h_2=0.25\pm0.15$ of the bachelor 
final state pion or kaon.    
The large uncertainties of the decay widths of the states $K^*_0(1430)$ and $K^*_0(1950)$ in the Table~\ref{tab1} result in quite 
small errors, which have been neglected, for these quasi-two-body predictions in the Tables~\ref{Res-1430} and \ref{Res-1950}.
The reason is that the variation effect of decay width $\Gamma$ in the denominator ${\mathcal{D}}_{K^*_0}$ of Eq.~(\ref{eq-vert-SKpi}) 
will be mainly canceled out by the uncertainty of $\Gamma_{{K^*_0}\to K\pi}$ 
(equals to $\Gamma\cdot {{\mathcal{B}}({K^*_0}\to K\pi)}$) in the numerator $g_{K_0^*K\pi}$.
For instance, the corresponding errors for the decay process $B^+\to  K_0^{*}(1430)^0 \pi^+\to K^+\pi^- \pi^+$ are 
$0.04\times 10^{-5}$ and $0.2\%$ for its branching fraction and direct $CP$ asymmetry, respectively, while for 
$B^+\to  K_0^{*}(1950)^0 \pi^+\to K^+\pi^- \pi^+$, the two errors are $0.01\times10^{-6}$ and $0.1\%$. 
There are other errors, which come from the uncertainties of the Wolfenstein parameters of the CKM matrix, the parameters 
in the distribution amplitudes for bachelor pion or kaon, 
the masses and the decay constants of the initial and final states, etc. are small and have been neglected. One can find that for 
those decay modes with the main contributions come from the annihilation diagrams of Fig.~\ref{fig-feyndiag}, their branching fraction 
errors generated from the variations of the $a_1$ and $a_3$ could be larger than the corresponding errors from $\omega_B$ or 
$\omega_{B_s}$, because there is no shape parameter for $B$ meson in the factorizable annihilation diagrams. 

\begin{figure}[tbp]   
\vspace{-0.3cm}
\centerline{\epsfxsize=8 cm \epsffile{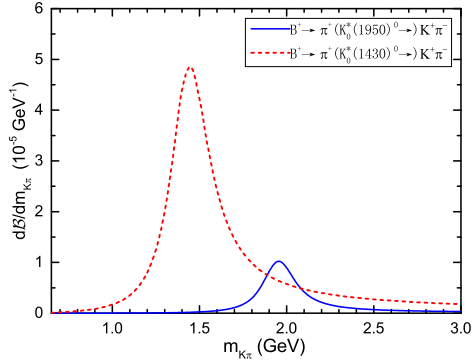}}
     \vspace{-0.5cm}
\caption{Differential branching fractions from threshold of $K\pi$ pair to $3$ GeV for the 
               $B^+\to K_0^{*}(1430)^0\pi^+\to K^+\pi^-\pi^+$ and $B^+\to K_0^{*}(1950)^0\pi^+\to K^+\pi^-\pi^+$ decays.}
\label{fig=depbr}
\end{figure}

In this work, the branching fraction of a quasi-two-body decay process involving the resonant state $K_0^{*}(1950)$ is predicted 
to be roughly one order smaller than the corresponding decay mode with the resonance $K_0^{*}(1430)$.
Or rather, the ratios between the $CP$ averaged branching fractions for the decays in Table~\ref{Res-1950} and Table~\ref{Res-1430}
with (without) the factorizable emission diagrams of Fig.~\ref{fig-feyndiag} (c) are about $12\%$-$15\%$ ($6\%$-$9\%$). 
The difference mainly originated from the $(S-P)(S+P)$ amplitude the Eq.~(\ref{exp-figc-Fsp}), which has the intermediate state 
invariant mass factor $m_B\sqrt\zeta$ ($\equiv\sqrt s$).  
This factor makes the proportion originated from Eq.~(\ref{exp-figc-Fsp}) in the total branching ratio
for a quasi-two-body decay mode invloving $K_0^{*}(1950)$ larger than that of the corresponding decay 
process including $K_0^{*}(1430)$ because of the larger pole mass of the resonance $K_0^{*}(1950)$. 
Take the decays $B^+\to  K_0^{*}(1430,1950)^0 \pi^+\to K^+\pi^- \pi^+$ as the examples, when we neglect the contribution from the 
$(S-P)(S+P)$ amplitude of the Eq.~(\ref{exp-figc-Fsp}), the ratio between two branching fractions of the decays 
$B^+\to  K_0^{*}(1950)^0 \pi^+\to K^+\pi^- \pi^+$ and $B^+\to  K_0^{*}(1430)^0 \pi^+\to K^+\pi^- \pi^+$ will drop to $8\%$ 
from about $15\%$. From the lines of the differential branching fractions for $B^+\to  K_0^{*}(1950)^0 \pi^+\to K^+\pi^- \pi^+$ 
and $B^+\to  K_0^{*}(1430)^0 \pi^+\to K^+\pi^- \pi^+$ in Fig.~\ref{fig=depbr}, one can find that the main 
portion of the branching fractions lies in the region around the corresponding pole mass of the intermediate states.

\begin{table}[thb] 
\begin{center}
\caption{PQCD predictions of the $CP$ averaged branching fractions and the direct $CP$
         asymmetries for the quasi-two-body $B\to K_0^*(1950)h \to K\pi h$ decays.}
\label{Res-1950}   
\begin{tabular}{l c l} \hline\hline
   \quad Decay modes       &    ~    &  \quad\; Quasi-two-body results               \\
\hline
  $B^+\to K_0^{*}(1950)^0 \pi^+\to K^+\pi^- \pi^+$    &${\mathcal B}(10^{-6})$
      & $3.36\pm0.86(\omega_B)\pm0.24(a_{3+1})\pm0.51(m^\pi_0{+}a^\pi_2)$    \\
              &  ${\mathcal A}_{CP}(\%)$
      & $1.5\pm0.3(\omega_B)\pm0.2(a_{3+1})\pm0.3(m^\pi_0{+}a^\pi_2)$    \\
   $B^+\to K_0^{*}(1950)^+ \pi^0\to K^0\pi^+ \pi^0$   &${\mathcal B}(10^{-6})$
      & $1.19\pm0.32(\omega_B)\pm0.08(a_{3+1})\pm0.21(m^\pi_0{+}a^\pi_2)$   \\
               &  ${\mathcal A}_{CP}(\%)$
      & $3.5\pm0.1(\omega_B)\pm0.4(a_{3+1})\pm0.2(m^\pi_0{+}a^\pi_2)$    \\
  $B^+\to K_0^{*}(1950)^+ \bar K^0\to K^0\pi^+ \bar K^0$     &${\mathcal B}(10^{-8})$
      & $1.86\pm0.04(\omega_B)\pm0.60(a_{3+1})\pm0.38(m^K_0{+}a^K_2)$  \\
              &  ${\mathcal A}_{CP}(\%)$
      & $-9.2\pm5.3(\omega_B)\pm4.0(a_{3+1})\pm2.8(m^K_0{+}a^K_2)$  \\
  $B^+\to \bar K_0^{*}(1950)^0 K^+\to K^-\pi^+ K^+$    &${\mathcal B}(10^{-7})$
      & $3.59\pm0.66(\omega_B)\pm0.54(a_{3+1})\pm0.54(m^K_0{+}a^K_2)$   \\
               &  ${\mathcal A}_{CP}(\%)$
      & $19.2\pm0.1(\omega_B)\pm7.4(a_{3+1})\pm1.4(m^K_0{+}a^K_2)$    \\
\hline
  $B^0\to K_0^{*}(1950)^+ \pi^-\to K^0\pi^+ \pi^-$ &${\mathcal B}(10^{-6})$
      & $2.99\pm0.77(\omega_B)\pm0.20(a_{3+1})\pm0.45(m^\pi_0{+}a^\pi_2)$    \\
              &  ${\mathcal A}_{CP}(\%)$
      & $1.9\pm0.5(\omega_B)\pm0.5(a_{3+1})\pm0.1(m^\pi_0{+}a^\pi_2)$    \\
  $B^0\to K_0^{*}(1950)^0 \pi^0\to K^+\pi^- \pi^0$ &${\mathcal B}(10^{-6})$
      & $2.01\pm0.50(\omega_B)\pm0.15(a_{3+1})\pm0.26(m^\pi_0{+}a^\pi_2)$    \\
              &  ${\mathcal A}_{CP}(\%)$
      & $0.4\pm0.6(\omega_B)\pm0.3(a_{3+1})\pm0.3(m^\pi_0{+}a^\pi_2)$   \\
  $B^0\to K_0^{*}(1950)^+K^-\to  K^0\pi^+ K^-$    &${\mathcal B}(10^{-9})$
      & $5.14\pm1.90(\omega_B)\pm1.66(a_{3+1})\pm0.29(m^K_0{+}a^K_2)$     \\
              &  ${\mathcal A}_{CP}(\%)$
      & $-2.8\pm10(\omega_B)\pm10.6(a_{3+1})\pm3.3(m^K_0{+}a^K_2)$    \\
  $B^0\to K_0^{*}(1950)^-K^+ \to \bar K^0\pi^- K^+$     &${\mathcal B}(10^{-8})$
      & $2.36\pm0.95(\omega_B)\pm1.10(a_{3+1})\pm0.06(m^K_0{+}a^K_2)$   \\
               &  ${\mathcal A}_{CP}(\%)$
      & $-1.0\pm2.4(\omega_B)\pm8.5(a_{3+1})\pm1.3(m^K_0{+}a^K_2)$    \\
  $B^0\to K_0^{*}(1950)^0 \bar K^0\to  K^+\pi^- \bar K^0$ &${\mathcal B}(10^{-8})$
      & $2.22\pm0.08(\omega_B)\pm1.05(a_{3+1})\pm0.35(m^K_0{+}a^K_2)$   \\
              &  ${\mathcal A}_{CP}(\%)$
      & \quad -    \\
  $B^0\to \bar K_0^{*}(1950)^0 K^0\to K^-\pi^+ K^0$ &${\mathcal B}(10^{-7})$
      & $3.36\pm0.64(\omega_B)\pm0.58(a_{3+1})\pm0.48(m^K_0{+}a^K_2)$   \\
              &  ${\mathcal A}_{CP}(\%)$
      & \quad -    \\
\hline
  $B_s^0\to K_0^{*}(1950)^- \pi^+ \to \bar K^0\pi^- \pi^+$ &${\mathcal B}(10^{-6})$
      & $3.35\pm0.59(\omega_B)\pm0.37(a_{3+1})\pm0.01(m^\pi_0{+}a^\pi_2)$    \\
              &  ${\mathcal A}_{CP}(\%)$
      & $12.9\pm7.0(\omega_B)\pm3.1(a_{3+1})\pm0.8(m^\pi_0{+}a^\pi_2)$   \\
  $B_s^0\to \bar K_0^{*}(1950)^0 \pi^0\to K^-\pi^+ \pi^0$ &${\mathcal B}(10^{-8})$
      & $3.74\pm0.35(\omega_B)\pm1.01(a_{3+1})\pm0.48(m^\pi_0{+}a^\pi_2)$    \\
              &  ${\mathcal A}_{CP}(\%)$
      & $57.1\pm4.0(\omega_B)\pm8.1(a_{3+1})\pm5.5(m^\pi_0{+}a^\pi_2)$   \\
  $B_s^0\to K_0^{*}(1950)^+ K^- \to K^0\pi^+ K^- $ &${\mathcal B}(10^{-6})$
      & $2.03\pm0.31(\omega_B)\pm0.19(a_{3+1})\pm0.32(m^K_0{+}a^K_2)$    \\
              &  ${\mathcal A}_{CP}(\%)$
      & $0.6\pm0.2(\omega_B)\pm1.0(a_{3+1})\pm0.9(m^K_0{+}a^K_2)$   \\
  $B_s^0\to K_0^{*}(1950)^- K^+ \to \bar K^0\pi^- K^+ $ &${\mathcal B}(10^{-6})~$
      & $1.26\pm0.15(\omega_B)\pm0.54(a_{3+1})\pm0.20(m^K_0{+}a^K_2)$  \\
              &  ${\mathcal A}_{CP}(\%)$
      & $-45.1\pm1.3(\omega_B)\pm4.6(a_{3+1})\pm5.7(m^K_0{+}a^K_2)$  \\
  $B_s^0\to  K_0^{*}(1950)^0 \bar K^0\to K^+\pi^- \bar K^0$ &${\mathcal B}(10^{-6})$
      & $2.13\pm0.33(\omega_B)\pm0.19(a_{3+1})\pm0.33(m^K_0{+}a^K_2)$  \\
              &  ${\mathcal A}_{CP}(\%)$
      & \quad -    \\
  $B_s^0\to \bar K_0^{*}(1950)^0 K^0\to K^-\pi^+ K^0$ &${\mathcal B}(10^{-7})$
      & $7.65\pm0.54(\omega_B)\pm4.56(a_{3+1})\pm1.48(m^K_0{+}a^K_2)$   \\
              &  ${\mathcal A}_{CP}(\%)$
      & \quad -    \\
\hline\hline
\end{tabular}
\end{center}
\end{table}

We must stress that the ratios between the corresponding branching fractions in Table~\ref{Res-1950} and 
Table~\ref{Res-1430}, and also the branching fractions in Table~\ref{Res-1950} for 
the quasi-two-body decays involving $K_0^{*}(1950)$ are squared dependent on the result 
$f_{K_0^*(1950)}m^2_{K_0^*(1950)}=0.0414$ GeV$^3$~\cite{prd63-074017}. If the value $0.0414$ becomes two times larger, 
 the ratios and the branching fractions in Table~\ref{Res-1950} will become four times larger than their current values.
In Ref.~\cite{epjc78-1019}, there are two branching fractions measured by LHCb to be
\begin{eqnarray}
{\mathcal B}(B^0\to \eta_c K^*_0(1950)^0 \to \eta_cK^+\pi^-)&=&(2.18\pm1.04\pm0.04^{+0.80}_{-1.43}\pm0.25)\times 10^{-5},\\
{\mathcal B}(B^0\to \eta_c K^*_0(1430)^0 \to \eta_cK^+\pi^-)&=&(14.50\pm2.10\pm0.28^{+2.01}_{-1.60}\pm1.67)\times 10^{-5}.
\end{eqnarray}
The two central values above give us the ratio about $0.15$ between these two branching factions, 
but there is no diagrams like Fig.~\ref{fig-feyndiag} (c) for $B^0\to \eta_c K^{*0}_0$ decays.
Because of the large errors for $B^0\to \eta_c K^*_0(1950)^0$, we can not extract the decay constant $f_{K^*_0(1950)}$ from this 
measurement. While from the data of the fit fractions for  $\eta_c\to K^0_SK^\pm\pi^\mp$ in~\cite{prd93-012005} and 
$\eta_c\to K^+K^-\pi^0$ in~\cite{prd89-112004} both from BaBar, one can expect a larger value than $0.0414$ GeV$^3$ for the 
$f_{K_0^*(1950)}m^2_{K_0^*(1950)}$.

\begin{table}[thb]  
\begin{center}
\caption{Comparison of the extracted predictions with the experimental measurements for the relevant two-body branching 
               fractions (in units of $10^{-6}$). The first error for the theoretical results is added in quadrature from the errors in 
               Table~\ref{Res-1430}, the second error comes from the uncertainty of 
               ${\mathcal B}(K^*_0(1430)\to K\pi)=0.93\pm0.10$~\cite{PDG-2018}.}
\label{tab-data}   
\begin{tabular}{l l l l} \hline\hline
   \;Two-body decays        & \;\;\, This work       & \;\;  Data             & \; Ref.       \\
\hline
  $B^+\to  K_0^{*}(1430)^0\pi^+$     &\;\; $36.6\pm11.3\pm3.9$\;       &\; $34.6\pm3.3\pm4.2^{+1.9}_{-1.8}$ 
       &\;BaBar~\cite{prd96-072001} \\
                                                           &                                       &\; $32.0\pm1.2\pm2.7^{+9.1}_{-1.4}\pm5.2$\;\;
       &\;BaBar~\cite{prd78-012004} \\
                                                           &                                       &\; $51.6\pm1.7\pm6.8^{+1.8}_{-3.1}$ 
       &\;\;Belle~~\cite{prl96-251803} \\
  $B^+\to K_0^{*}(1430)^+ \pi^0$      &\;\; $12.7\pm4.2\pm1.4$\;         &\; $11.9\pm1.7\pm1.0^{+0.0}_{-1.3}$      
      &\;BaBar~\cite{prd96-072001}   \\
  $B^0\to K_0^{*}(1430)^+ \pi^-$       &\;\; $33.4\pm10.2\pm3.6$\;       &\;  $29.9^{+2.3}_{-1.7}\pm1.6\pm0.6\pm3.2$\;\;    
      &\;BaBar~\cite{prd80-112001}   \\
                                                            &                                       &\; $49.7\pm3.8\pm6.7^{+1.2}_{-4.8}$ 
      &\;\;Belle~~\cite{prd75-012006} \\
\hline\hline
\end{tabular}
\end{center}
\end{table}

The two-body branching fractions for $B\to K^*_0 h$ can be extracted from the quasi-two-body predictions 
of this work with the relation
\begin{eqnarray}
\Gamma(B\to K^*_0 h\to K\pi h)=\Gamma(B\to K^*_0 h)\times{\mathcal B}(K^*_0 \to K\pi).
\label{def-fac-rel}
\end{eqnarray}
In Ref.~\cite{prd67-054021}, a parameter $\eta$ was defined to measure the violation of the factorization relation the 
Eq.~(\ref{def-fac-rel}) in the $D$ meson decays. For the $B\to K_0^{*}(1430) h$ and $B\to K_0^{*}(1430) h\to K\pi h$ decays, 
we have %
\begin{eqnarray}
\eta&=&\frac{\Gamma(B\to K_0^{*}(1430)h\to K\pi h)}{\Gamma(B\to K_0^{*}(1430)h)\times{\mathcal B}(K_0^{*}\to K\pi)}\nonumber\\
      &\approx&\frac{m^2_{K_0^{*}(1430)}}{4\pi m_{B}}\frac{\Gamma_{K_0^{*}(1430)}}{\hat q_hq_0}
             \int^{(m_{B}-m_h)^2}_{(m_K+m_\pi)^2}\frac{ds}{s}\frac{\lambda^{1/2}(m^2_B,s,m^2_h)\lambda^{1/2}(s,m^2_K,m^2_\pi) }
             {(s-m^2_{K_0^{*}(1430)})^2+(m_{K_0^{*}(1430)}\Gamma_{K^*_0}(s))^2}, \;\;
\label{def-parm-eta}
\end{eqnarray}
where $\lambda(a,b,c)=a^2+b^2+c^2-2ab-2ac-2bc$, the $\hat q_h$ is the expression of Eq.~(\ref{def-qh}) in the rest frame 
of $B$ meson and fixed at $s=m^2_{K_0^{*}(1430)}$.  With Eq.~(\ref{def-parm-eta}), we have $\eta=0.90$ for the decays 
$B^+\to  K_0^{*}(1430)^0\pi^+$,  
which means the violation of the factorization relation is not large when neglecting the effect of the invariant mass $s$ in the 
decay amplitudes of the quasi-two-body decays. In order to check this conclusion, we calculate the decay 
$B^+\to  K_0^{*}(1430)^0\pi^+$ in the two-body framework of the PQCD approach, and we have 
${\mathcal B}(B^+\to  K_0^{*}(1430)^0\pi^+)=35.2\times 10^{-6}$, which is about $96.2\%$ of the result in Table~\ref{tab-data} extracted with Eq~(\ref{def-fac-rel}), and ${\mathcal A}_{CP}(B^+\to  K_0^{*}(1430)^0\pi^+)=-1.0\%$ is consistent with the 
$-1.3\%$ in Table~\ref{Res-1430}.

The comparison of the PQCD branching fractions with the experimental measurements for the two-body decays 
$B^+\to  K_0^{*}(1430)^0\pi^+$,  $B^+\to K_0^{*}(1430)^+ \pi^0$ and $B^0\to K_0^{*}(1430)^+ \pi^-$ are shown in the 
Table~\ref{tab-data}, with the first error added in quadrature from the errors in Table~\ref{Res-1430} and the second error 
comes from the uncertainty of ${\mathcal B}(K^*_0(1430)\to K\pi)=0.93\pm0.10$~\cite{PDG-2018} for these theoretical results. 
The branching fraction and direct $CP$ asymmetry for $B^+\to  K_0^{*}(1430)^0\pi^+$ in {\it Review of Particle 
Physics}~\cite{PDG-2018} averaged from the results in~\cite{prd96-072001,prd78-012004,prl96-251803} are 
$39^{+6}_{-5}\times 10^{-6}$ and $0.061\pm0.032$, respectively, which are consistent with the predictions 
$(36.6\pm11.3\pm3.9)\times 10^{-6}$ in Table~\ref{tab-data} and $(-1.3\pm0.5)\%$ in Table~\ref{Res-1430}.
Because of the large uncertainty of the ${\mathcal A}_{CP}=0.26^{+0.18}_{-0.14}$ for $B^+\to K_0^{*}(1430)^+ \pi^0$ 
in~\cite{PDG-2018}, we can not evaluate the significance of the prediction $(1.5\pm1.0)\%$, but our branching fraction 
agrees very well with BaBar's result in~\cite{prd96-072001} for this decay mode. For the decay $B^0\to K_0^{*}(1430)^+ \pi^-$, 
one has two results as listed in Table~\ref{tab-data} from BaBar and Belle Collaborations, its average ${\mathcal B}$ is presented 
to be $(33\pm7)\times 10^{-6}$ in {\it Review of Particle Physics}~\cite{PDG-2018}, this value agrees well with the PQCD prediction 
$(33.4\pm10.2\pm3.6)\times 10^{-6}$. There is an upper limit of $2.2\times 10^{-6}$ for the decay $B^+\to \bar K^*_0(1430)^0K^+$, 
which is below our expectation. Our predictions in this work will be tested by future experiments.
In the very recent work, LHCb Collaboration presented the branching fractions for the combined decays 
$B^0_s\to$~\KorKbar\hspace{0.2truemm}$^0 \pi^\pm K^\mp$ as~\cite{jhep1906-114}
\begin{eqnarray}%
{\mathcal B}(B^0_s\to K^*_0(1430)^\pm K^\mp\to \text{\KorKbar}\hspace{0.1truemm}^0\pi^\pm K^\mp)
    \! =\!(19.4\pm1.4\pm0.4\pm15.6\pm2.0\pm0.3)\times10^{-6}, \quad\, \\
{\mathcal B}(B^0_s\to \text{\KorKbar}\hspace{0.1truemm}^*_0(1430)^0 \text{\KorKbar}\hspace{0.1truemm}^0   
      \to K^\mp\pi^\pm  \text{\KorKbar}\hspace{0.1truemm}^0)
     \! =\!(20.5\pm1.6\pm0.6\pm5.7\pm2.2\pm0.3)\times10^{-6}, \quad\;
\end{eqnarray}
which are in agreement with the PQCD predictions in Table~\ref{Res-Bs-com}.

\begin{table}[thb] 
\begin{center}
\caption{PQCD predictions of the $CP$ averaged branching fractions and the direct $CP$ asymmetries for the combined decays
              $B^0_s\to K\pi K$, with the resonances $K^*_0(1430)$ and   $K^*_0(1950)$ as the intermediate states.}
\label{Res-Bs-com}   
\begin{tabular}{l c l} \hline\hline
  \quad Decay modes       &    ~    &  \quad\; Quasi-two-body results    \\     \hline  \vspace{-0.35cm}\\
   $B^0_s\to K^*_0(1430)^\pm K^\mp\to\text{\KorKbar}\hspace{0.1truemm}^0\pi^\pm K^\mp$
     &${\mathcal B}(10^{-5})$
          &$1.97\pm0.45(\omega_B)\pm0.10(a_{3+1})\pm0.43(m^K_0{+}a^K_2)$   \\
     &${\mathcal A}_{CP}(\%)$
          &$-7.7\pm1.5(\omega_B)\pm1.3(a_{3+1})\pm4.3(m^K_0{+}a^K_2)$   \\
  $B^0_s\to \text{\KorKbar}\hspace{0.1truemm}^*_0(1430)^0 \text{\KorKbar}\hspace{0.1truemm}^0   
      \to K^\mp\pi^\pm  \text{\KorKbar}\hspace{0.1truemm}^0$
      &${\mathcal B}(10^{-5})$
          &$1.50\pm0.36(\omega_B)\pm0.09(a_{3+1})\pm0.40(m^K_0{+}a^K_2)$   \\
      &${\mathcal A}_{CP}(\%)$
          &\;\quad -\;   \\
  $B^0_s\to K^*_0(1950)^\pm K^\mp\to \text{\KorKbar}\hspace{0.1truemm}^0\pi^\pm K^\mp$
      &${\mathcal B}(10^{-6})$
          &$3.20\pm0.69(\omega_B)\pm0.21(a_{3+1})\pm0.62(m^K_0{+}a^K_2)$    \\
     &${\mathcal A}_{CP}(\%)$
          &$3.1\pm0.2(\omega_B)\pm2.9(a_{3+1})\pm1.6(m^K_0{+}a^K_2)$   \\
  $B^0_s\to \text{\KorKbar}\hspace{0.1truemm}^*_0(1950)^0  \text{\KorKbar}\hspace{0.1truemm}^0   
      \to K^\mp\pi^\pm  \text{\KorKbar}\hspace{0.1truemm}^0$
     &${\mathcal B}(10^{-6})$
          &$2.67\pm0.59(\omega_B)\pm0.25(a_{3+1})\pm0.61(m^K_0{+}a^K_2)$   \\
     &${\mathcal A}_{CP}(\%)$
          &\;\quad -\;   \\
\hline\hline
\end{tabular}
\end{center}
\end{table}

On the experimental side, the LASS parametrization~\cite{npb296-493,prd72-072003} 
\begin{eqnarray}
R(s)=\frac{\sqrt s}{q{\rm cot}\delta_B-iq}
      +e^{2i\delta_B}\frac{m_{0}\Gamma_0\frac{m_0}{q_0}}{m^2_0-s-im_0\Gamma_0\frac{q}{m}\frac{m_0}{q_0}}\;,
\label{fun-LASS}
\end{eqnarray}
are employed in most cases to describe the $S$-wave $K\pi$ system, where $m_0$ and $\Gamma_0$ are now the pole mass 
and full width for $K^*_0(1430)$, and ${\rm cot}\delta_B=\frac{1}{aq}+\frac12 rq$ with the parameters $a=2.07\pm0.10$ GeV$^{-1}$ 
and $r=3.32\pm0.34$ GeV$^{-1}$~\cite{prd72-072003}. The relativistic Breit-Wigner term of Eq.~(\ref{fun-LASS}) is different from 
Eq.~(\ref{eq-vert-SKpi}). Before the $F_{K\pi}(s)$ in Eqs.~(\ref{def-wavefun-twist2})-(\ref{def-wavefun-twist3}) be replaced by 
the LASS expression, a coefficient is needed for $R(s)$. We have the replacement
\begin{eqnarray}
F_{K\pi}(s)\to     
           \hat R(s)= \frac{q_0}{m^2_0\Gamma_0}g_{K_0^*(1430)K\pi}\bar f_{K^*_0(1430)} R(s)
\end{eqnarray}
on the theoretical side. 
With $\hat R(s)$ in the concerned quasi-two-body decay amplitudes, one could in principle have the predictions for the 
decays $B\to (K\pi)^*_0 h$, including the results as same as the values in the Table~\ref{Res-1430} for the resonance $K^*_0(1430)$ 
and the contributions from the nonresonant effective range term. 
But we argue that, considering the nonresonant term of a three-body decay amplitude should not be included 
in the resonance distribution amplitudes the Eq.~(\ref{def-wavefun-Kpi}), 
it's improper for the effective range term of the Eq.~(\ref{fun-LASS}) to be studied in the 
quasi-two-body framework with the same expressions of the decay amplitudes in Appendix~\ref{sec-appx-a}.

The two-body decays $B^+\to  K_0^{*}(1430)^0\pi^+$,  $B^+\to K_0^{*}(1430)^+ \pi^0$, $B^0\to K_0^{*}(1430)^+ \pi^-$ and 
$B^0\to K_0^{*}(1430)^0 \pi^0$ have been studied in Ref.~\cite{prd73-014017} and updated in~\cite{prd87-114001} in the QCDF 
with $K_0^{*}(1430)$ being the first excited states of $K_0^{*}(700)$ (scenario $1$) or the lowest lying scalar state (scenario $2$), 
and in the scenario $2$ $K_0^{*}(700)$ is treated as a four-quark state. In view of the discussions for $K_0^{*}(700)$ 
in~\cite{prd15-267,prl48-659,prd41-2236,prl93-212002,pr389-61,pr409-1,prd74-014028}, we will consider only the results for 
the $K_0^{*}(1430)$ in the scenario $2$ in this work. The  branching fractions in~\cite{prd73-014017,prd87-114001} 
for the four decays invloving the $K_0^{*}(1430)$ are all smaller when comparing with the measurements and our results but with 
quite large errors as shown in Table~\ref{Res-literature}. The difference between our predictions and the concerned results 
in~\cite{prd73-014017,prd87-114001} may be partly due to the dynamical enhancement of penguin contributions in PQCD approach 
as discussed in detail in~\cite{plb504-6,prd63-054008,prd74-094020},  and also due to the small values for the decay constant 
$f_{K_0^{*}(1430)}$ and the $B\to K_0^{*}(1430)$ and $B\to \pi$ transition form factors $F^{BK_0^{*}}_{0,1}$ and  $F^{B\pi}_{0,1}$ 
in~\cite{prd73-014017,prd87-114001}. The value $f_{K_0^{*}(1430)}=34$ MeV~\cite{prd73-014017} will make our branching ratios 
involving $K_0^{*}(1430)$ $1.5$ times smaller than the results in Table~\ref{Res-literature}.
With the parameters in this work, the form factors $F^{B\pi}_{0,1}(0)=0.26$ and $F^{BK_0^{*}(1430)}_{0,1}(0)=0.42$ could be induced 
in the PQCD approach. The value $0.26$ is close to $0.25$ for $F^{B\pi}_{0,1}(0)$ in~\cite{prd73-014017,prd87-114001}, 
but the $0.42$ for $F^{BK_0^{*}}_{0,1}(0)$ is two times larger than the value $0.21$ in Refs.~\cite{prd73-014017,prd87-114001}.
In the PQCD approach, the two-body decays $B\to K^*_0(1430)\pi$ were 
studied in~\cite{epjc50-877}, with the branching fractions larger than the corresponding results of this work except the 
decay $B^0\to K_0^{*}(1430)^0 \pi^0$ which is $18.4^{+4.4+1.5+4.0}_{-3.9-1.4-2.9}\times 10^{-6}$ in~\cite{epjc50-877} as listed 
in Table~\ref{Res-literature}. The result $28.8^{+6.8+1.9+3.2}_{-6.1-1.9-3.5}\times 10^{-6}$ in~\cite{epjc50-877} is about 
double of our prediction and BaBar's measurement~\cite{prd96-072001} for the decay $B^+\to K_0^{*}(1430)^+ \pi^0$.
The difference between the results in~\cite{epjc50-877} and our predictions could be be explained as the different input parameters. 
The decays $B\to K^*_0(1430) K$ have been studied in the QCDF in~\cite{prd91-074022}. 
One can find the comparison of relevant branching fractions in Table~\ref{Res-lit-KK}. 
The ${\mathcal A}_{CP}=-22.51^{+4.90+5.63+19.61}_{-7.57-9.36-22.86}\%$ for the decay $B^+\to K_0^{*}(1430)^+ \bar K^0$ 
in~\cite{prd91-074022} is consistent with the result $(-18.4\pm5.8\pm2.7\pm5.4)\%$ in Table~\ref{Res-1430}, while the 
${\mathcal A}_{CP}=-2.60^{+1.61+0.59+3.52}_{-1.76-0.59-5.47}\%$ for $B^+\to \bar K_0^{*}(1430)^0 K^+$ in~\cite{prd91-074022}  
is smaller than the PQCD prediction $(17.9\pm0.4\pm8.0\pm0.9)\%$ in this work and with an opposite sign.

\begin{table}[thb]  
 \begin{center}
\caption{Comparison of the extracted predictions with the results in literature for the relevant two-body 
               branching fractions (in units of $10^{-6}$). The sources of the errors of our results are the same as 
               in Table~\ref{tab-data}.}
\label{Res-literature}   
\begin{tabular}{l l l l} \hline\hline
   \;Two-body decays        & \;\;\, This work       & \;\;  Theory            & \;\,Ref.      \\
\hline
  $B^+\to  K_0^{*}(1430)^0\pi^+$\;     &\;\; $36.6\pm11.3\pm3.9$\;\;\;       &\; $11.0^{+10.3+7.5+49.9}_{-6.0-3.5-10.1}$\;\;
       &\;\cite{prd73-014017} \\
                                                           &                                       &\; $12.9^{+4.6+4.1+38.5}_{-3.7-3.4-9.1}$\;\;
       &\;\cite{prd87-114001} \\
                                                           &                                       &\; $47.6^{+11.3+3.7+6.9}_{-10.1-3.6-5.1}$ 
       &\;\cite{epjc50-877} \\
  $B^+\to K_0^{*}(1430)^+ \pi^0$      &\;\; $12.7\pm4.2\pm1.4$\;         &\; $5.3^{+4.7+1.6+22.3}_{-2.8-1.7-4.7}$      
      &\;\cite{prd73-014017}   \\
                                                           &                                       &\; $7.4^{+2.4+2.1+20.1}_{-1.9-1.8-5.0}$\;\;
       &\;\cite{prd87-114001} \\
                                                           &                                       &\; $28.8^{+6.8+1.9+3.2}_{-6.1-1.9-3.5}$ 
       &\;\cite{epjc50-877} \\
  $B^0\to K_0^{*}(1430)^+ \pi^-$       &\;\; $33.4\pm10.2\pm3.6$\;       &\;  $11.3^{+9.4+3.7+45.8}_{-5.8-3.7-9.9}$\;\;    
      &\;\cite{prd73-014017}   \\
                                                            &                                       &\; $13.8^{+4.5+4.1+38.3}_{-3.6-3.5-9.5}$ 
      &\;\cite{prd87-114001} \\
                                                           &                                       &\; $43.0^{+10.2+3.1+7.0}_{-9.1-2.9-5.2}$ 
       &\;\cite{epjc50-877} \\
 $B^0\to K_0^{*}(1430)^0 \pi^0$       &\;\; $22.4\pm6.6\pm2.4$\;       &\;  $6.4^{+5.4+2.2+26.1}_{-3.3-2.1-5.7}$\;\;    
      &\;\cite{prd73-014017}   \\
                                                            &                                       &\; $5.6^{+2.6+2.4+18.8}_{-1.3-1.2-3.9}$ 
      &\;\cite{prd87-114001} \\
                                                           &                                       &\; $18.4^{+4.4+1.5+4.0}_{-3.9-1.4-2.9}$ 
       &\;\cite{epjc50-877} \\
\hline\hline
\end{tabular}
\end{center}
\end{table}

\begin{table}[thb]  
\begin{center}
\caption{Comparison of the extracted predictions with the QCDF results in~\cite{prd91-074022} for the relevant two-body 
               branching fractions (in units of $10^{-7}$). The sources of the errors of our results are the same as 
               in Table~\ref{tab-data}.}
\label{Res-lit-KK}   
\begin{tabular}{l l l} \hline\hline
   \;Two-body decays        & \;\;\, This work       & \;\;  QCDF~\cite{prd91-074022}              \\
\hline
 $B^+\to K_0^{*}(1430)^+ \bar K^0$      &\;\; $3.76\pm2.16\pm0.40$\;\;\;    &\;  $1.14^{+0.54+1.40+1.17}_{-0.38-0.56-0.92}$  \\
 $B^+\to \bar K_0^{*}(1430)^0 K^+$ \;   &\;\; $39.9\pm13.8\pm4.3$\;\;        &\;  $33.70^{+10.33+5.52+3.37}_{-8.47-4.82-3.94}$  \\
 $B^0\to K_0^{*}(1430)^+  K^-$              &\;\; $0.93\pm0.61\pm0.10$\;        &\;  $1.07^{+0.72+0.03+2.27}_{-0.47-0.04-0.97}$  \\
 $B^0\to K_0^{*}(1430)^- K^+$               &\;\; $6.19\pm3.95\pm0.67$\;        &\;  $0.58^{+0.45+0.02+0.14}_{-0.29-0.03-0.05}$  \\
 $B^0\to  K_0^{*}(1430)^0 \bar K^0$      &\;\; $4.90\pm3.34\pm0.53$\;\;      &\;  $2.39^{+1.20+1.95+2.67}_{-0.85-0.90-2.00}$  \\
 $B^0\to \bar K_0^{*}(1430)^0 K^0$       &\;\; $46.1\pm15.0\pm5.0$\;          &\; $40.47^{+13.36+6.09+6.06}_{-10.77-5.38-6.16}$  \\
\hline\hline
\end{tabular}
\end{center}
\end{table}

With $m_{K\pi}$ in the region $(0.64\!\sim\!1.76)$ GeV, 
the branching ratios of the decay processes $B^-\to [\bar K^*_0(1430)^0\to K^-\pi^+]\pi^-$ and 
$\bar B^0 \to [K^*_0(1430)^-\to \bar K^0\pi^-]\pi^+$ were calculated in QCDF in Ref.~\cite{prd79-094005} with the predictions 
$(11.6\pm0.6)\times 10^{-6}$ and $(11.1\pm0.5)\times 10^{-6}$, respectively. 
These two decays have also been studied in Ref.~\cite{prd81-094033}  in the $m_{K\pi}$ region $(1.0\sim1.76)$ GeV
and the branching ratios are $(12.11\pm0.32)\times 10^{-6}$ and $(11.05\pm0.25)\times 10^{-6}$, respectively. 
In the PQCD approach we have $(16.6\pm5.3)\times 10^{-6}$ and $(15.2\pm4.7)\times 10^{-6}$ in the region 
$m_{K\pi}\!\in\!(0.64\sim1.76)$ GeV,  $(16.4\pm5.1)\times 10^{-6}$ and $(15.0\pm4.6)\times 10^{-6}$ in the region 
$m_{K\pi}\!\in\!(1.0\sim1.76)$ GeV for the branching ratios of the decays $B^+\to  K_0^{*}(1430)^0 \pi^+\to K^+\pi^- \pi^+$ and 
$B^0\to K_0^{*}(1430)^+ \pi^-\to K^0\pi^+ \pi^-$, respectively, which are consistent with the results in 
Refs.~\cite{prd79-094005,prd81-094033} within errors. The three-body decays $B\to K\pi h$ have been discussed in detail 
in Refs.~\cite{prd88-114014,prd94-094015} in QCDF. The comparison of PQCD predictions in this work with the related results 
in~\cite{prd88-114014,prd94-094015} are listed in Table~\ref{Res-theory}. From Table~\ref{tab-data} and Table~\ref{Res-theory}, 
one can find that the PQCD predictions are totally larger than the QCDF results~\cite{prd88-114014,prd94-094015} but closer to 
the available data.

\begin{table}[thb]  
\begin{center}
\caption{Comparison of the PQCD predictions with the theoretical results for the relevant quasi-two-body 
               branching fractions (in units of $10^{-6}$). The errors of this work have been added in quadrature.}
\label{Res-theory}   
\begin{tabular}{l l l c} \hline\hline
   \;\; Decay modes        & \;\; This work       & \;\,  Theory            & \,Ref.       \\
\hline
  $B^+\to  K_0^{*}(1430)^0 \pi^+\to K^+\pi^- \pi^+$     &\;\; $22.7\pm7.0$\;       &\; $11.3^{+0.0+3.3+0.1}_{-0.0-2.8-0.1}$\;\,
       &\;\cite{prd88-114014}\\
                                                           &                                       &\; $11.5^{+0.0+3.3+0.0}_{-0.0-2.8-0.0}$\;\;
       &\;\cite{prd94-094015} \\
  $B^+\to K_0^{*}(1430)^+ \pi^0\to K^0\pi^+ \pi^0$     &\;\; $7.86\pm2.61$\;\;\,     &\; $5.4^{+0.0+1.6+0.1}_{-0.0-1.4-0.1}$ 
       &\;\cite{prd88-114014}\\
                                                           &                                       &\; $5.6^{+0.0+1.6+0.0}_{-0.0-1.4-0.0}$\;\;
       &\;\cite{prd94-094015} \\
  $B^+\to \bar K_0^{*}(1430)^0 K^+\to K^-\pi^+ K^+$     &\;\; $2.86\pm0.85$\;       &\; $1.0^{+0.0+0.2+0.0}_{-0.0-0.2-0.0}$ 
       &\;\cite{prd88-114014}\\
                                                           &                                       &\; $1.0^{+0.0+0.2+0.0}_{-0.0-0.2-0.0}$\;\;
       &\;\cite{prd94-094015} \\
  $B^0\to K_0^{*}(1430)^+ \pi^-\to K^0\pi^+ \pi^-$     &\;\; $20.7\pm6.3$\;       &\; $10.3^{+0.0+2.9+0.0}_{-0.0-2.5-0.0}$ 
       &\;\cite{prd88-114014}\\
                                                           &                                       &\; $10.6^{+0.0+3.0+0.0}_{-0.0-2.6-0.0}$\;\;
       &\;\cite{prd94-094015} \\
  $B^0\to  K_0^{*}(1430)^0 \pi^0\to K^+\pi^- \pi^0$     &\;\; $13.9\pm4.1$\;       &\; $4.1^{+0.0+1.4+0.0}_{-0.0-1.2-0.0}$ 
       &\;\cite{prd88-114014}\\
                                                           &                                       &\; $4.2^{+0.0+1.4+0.0}_{-0.0-1.2-0.0}$\;\;
       &\;\cite{prd94-094015} \\
\hline\hline
\end{tabular}
\end{center}
\end{table}

\begin{figure}[H]   
\vspace{-0.3cm}
\centerline{\epsfxsize=8 cm \epsffile{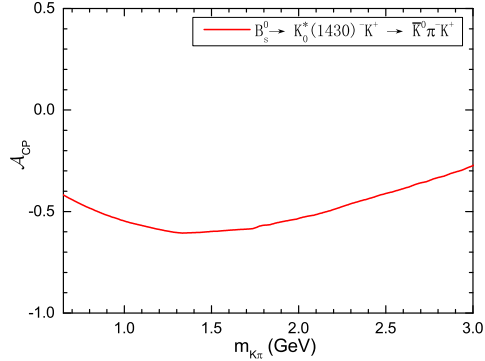}}
     \vspace{-0.5cm}
\caption{Differential direct $CP$ asymmetry 
               for the decay $B_s^0\to K_0^{*}(1430)^- K^+\to \bar K^0\pi^- K^+$.}
\label{fig=depAcp}
\end{figure}

There is no direct $CP$ asymmetries for $B^0_{(s)}\to K^{*0}_0 \bar K^0$ and $B^0_{(s)}\to\bar K^{*0}_0 K^0$ in 
Tables~\ref{Res-1430}, \ref{Res-1950}, because these decays have contributions only from the penguin operators in 
their decay amplitudes. For the decays $B^0\to K_0^{*}(1430)^+ \pi^-\to K^0\pi^+ \pi^-$ and                      
$B^0\to  K_0^{*}(1430)^0 \pi^0\to K^+\pi^- \pi^0$ via the $b\to s q\bar q$ transition at quark level, the very small proportion  
of the total branching ratio from the current-current operators led to the small direct $CP$ asymmetries 
for these two decays as shown in Table~\ref{Res-1430}. The same pattern will appear again for the decays 
$B^+\to  K_0^{*}(1430)^0 \pi^+\to K^+\pi^- \pi^+$ and $B^+\to K_0^{*}(1430)^+ \pi^0\to K^0\pi^+ \pi^0$, and also for the 
corresponding decays with the $K_0^{*}(1430)$ be  replaced by the $K_0^{*}(1950)$ as the intermediate, but not for the decays 
$B_s^0\to  K_0^{*}(1430)^-  \pi^+\to \bar K^0\pi^- \pi^+$ and $B_s^0\to \bar K_0^{*}(1430)^0 \pi^0\to K^-\pi^+ \pi^0$
via the $b\to d q\bar q$ transition. The interference between the weak and the strong phases of the decay amplitudes 
from current-current 
and penguin operators results in the large direct $CP$ asymmetries for the $B_s^0\to  K_0^{*}(1430)^-  \pi^+\to \bar K^0\pi^- \pi^+$ 
and $B_s^0\to \bar K_0^{*}(1430)^0 \pi^0\to K^-\pi^+ \pi^0$ decays. As an example, we display the differential distribution curve 
of the ${\mathcal A}_{CP}$ in $m_{K\pi}$ for the decay process $B_s^0\to K_0^{*}(1430)^- K^+\to \bar K^0\pi^- K^+$ in 
Fig.~\ref{fig=depAcp}.  

For the decays $B^+\to  K_0^{*}(1430)^0\pi^+$ and $B^0\to K_0^{*}(1430)^+ \pi^-$, with the isospin limit, 
one has the ratio~\cite{epjc50-877}
\begin{eqnarray}
R=\frac{\tau_{B^0}}{\tau_{B^+}}\frac{{\mathcal B}(B^+\to  K_0^{*}(1430)^0\pi^+)}{{\mathcal B}(B^0\to K_0^{*}(1430)^+ \pi^-)}
      \approx 1\,.
\end{eqnarray}
With the predictions in Table~\ref{Res-literature}, we have the ratio $R=1.017\pm0.003$ in this work. The small error for $R$ 
is because the cancellation between the errors of two branching ratios, which means the increase or the decrease of the parameters 
that caused the errors will result in nearly identical change of the weight for the numerator and denominator of $R$.
For the decays $B^+\to  K_0^{*}(1430)^+\pi^0$ and $B^0\to K_0^{*}(1430)^0 \pi^0$, the diagrams of 
Fig.~\ref{fig-feyndiag} (a), (c), (d) will contribute to the branching fractions, the decay amplitudes from Fig.~\ref{fig-feyndiag} (a) are same 
for both $B^+\to  K_0^{*}(1430)^+\pi^0$ and $B^0\to K_0^{*}(1430)^0 \pi^0$, but the decay amplitudes from 
Fig.~\ref{fig-feyndiag} (c), (d) have the opposite sign considering the difference for $\bar u u$ and $\bar d d$ to form a neutral pion.
It is not strange for the ratio between the branching fractions of $B^+\to  K_0^{*}(1430)^+\pi^0$ and $B^0\to K_0^{*}(1430)^0 \pi^0$ 
away from unity.

A relation for the direct $CP$ asymmetries of the two-body decays $B^+\to K^+\pi^0,\, B^+\to K^0\pi^+,\, B^0\to K^+\pi^-$ and 
$B^0\to K^0\pi^0$ was suggested in Ref.~\cite{plb627-82} as
\begin{eqnarray}
\label{relation-kpi}
{\mathcal A}_{CP}(B^+ &\to & K^+\pi^0)
     \frac{2{\mathcal B}(B^+\to K^+\pi^0)}{{\mathcal B}(B^0\to K^+\pi^-)}\frac{\tau_{B^0}}{\tau_{B^+}}  
     +{\mathcal A}_{CP}(B^0\to K^0\pi^0)  \frac{2{\mathcal B}(B^0\to K^0\pi^0)}{{\mathcal B}(B^0\to K^+\pi^-)} \nonumber\\     
&=&{\mathcal A}_{CP}(B^0\to K^+\pi^-) + {\mathcal A}_{CP}(B^+\to K^0\pi^+) 
    \frac{{\mathcal B}(B^+\to K^0\pi^+)}{{\mathcal B}(B^0\to K^+\pi^-)}\frac{\tau_{B^0}}{\tau_{B^+}}.\;
\end{eqnarray}
Considering the same transitions at quark level, one could extend the Eq.~(\ref{relation-kpi}) to the $B\to K^*_0(1430)\pi$ decays with 
the replacement $K\to K^*_0(1430)$. This relation is satisfied within errors with the 
${\mathcal A}_{CP}(B^0 \to K^*_0(1430)^+\pi^-)=(0.3\pm0.9)\%$, ${\mathcal A}_{CP}(B^+\to K^*_0(1430)^0\pi^+)=(-1.3\pm0.5)\%$,  
${\mathcal A}_{CP}(B^+\to K^*_0(1430)^+\pi^0)=(1.5\pm1.0)\%$ and ${\mathcal A}_{CP}(B^0\to K^*_0(1430)^0\pi^0)=(-1.8\pm0.5)\%$, 
and relevant branching fractions in Table~\ref{Res-1430}. One can find that the relation Eq.~(\ref{relation-kpi}) will also hold for 
$B\to K^*_0(1950)\pi$ decays with the values in Table~\ref{Res-1950}.
\section{Conclusion}
In this work, we studied the contributions from the resonant state $K_0^{*}(1430)$ and, for the first time, 
from the resonance $K_0^{*}(1950)$ in the three-body decays $B\to K\pi h$ in the PQCD approach. 
The crucial nonperturbative input factor $F_{K\pi}(s)$ in the distribution amplitudes of the $S$-wave 
$K\pi$ system was derived from the matrix element of the vacuum to $K\pi$ final state and was related to the scalar time-like form 
factor $F_0^{K\pi}(s)$ by the relation {$F_{K\pi}(s)={B_0 }/{m_{K_0^*}} F_0^{K\pi}(s)$}. This relation also means that the LASS 
parametrization for the $(K\pi)^*_0$ system which frequently appeared in the experimental works cannot be adopted directly for the 
$K\pi$ system distribution amplitudes in the PQCD approach. 

With $f_{K_0^*(1430)}m^2_{K_0^*(1430)}=0.0842\pm0.0045$ GeV$^3$ and $f_{K_0^*(1950)}m^2_{K_0^*(1950)}=0.0414$ GeV$^3$, 
the branching fractions and the direct $CP$ asymmetries for the concerned quasi-two-body decays 
$B\to K^*_0(1430,1950)h\to K\pi h$ were calculated. 
An important conclusion is that the $CP$ averaged  branching fraction of a quasi-two-body process with  
$K_0^{*}(1950)$ as the intermediate state is about one order smaller than the corresponding decay mode involving the resonance $K_0^{*}(1430)$.  In view of the important contribution from the $S$-wave $K\pi$ system for the $B\to K\pi h$ decays, it is not 
appropriate to neglect the $K_0^{*}(1950)$ in the theoretical or experimental studies for the relevant three-body $B$ meson decays.
We compared our predictions with the related results in literature and found the predictions in this work for the relevant 
decays agree well with the existing experimental results from BaBar, Belle and LHCb Collaborations.

\begin{acknowledgments}
This work was supported in part by the National Natural Science Foundation of China 
under Grants No.~11547038, No.~11575110 and No.~11947011; 
by the Natural Science Foundation of Jiangsu Province under Grant No. BK20191010.
\end{acknowledgments}                 

\appendix
\section{Decay amplitudes}\label{sec-appx-a}

The Lorentz invariant decay amplitude ${\mathcal A}$ for the quasi-two-body decay $B\to K^{*}_0 h\to K\pi h$ in 
the PQCD approach, according to Fig.~\ref{fig-feyndiag},  is given by~\cite{plb561-258,plb763-29}
\begin{eqnarray}
{\mathcal A}=\Phi_B\otimes H\otimes\Phi_{h}\otimes \Phi_{K\pi}\;.
\end{eqnarray}
The symbol $\otimes$ here means convolutions in parton momenta, the hard kernel $H$ contains one hard gluon exchange 
at the leading order in strong coupling $\alpha_s$ as in the two-body formalism. The distribution amplitudes 
$\Phi_B, \Phi_{h}$ and $\Phi_{K\pi}$ absorb the nonperturbative dynamics in the relevant decay processes. 

The $B$ meson light-cone matrix element can be decomposed as~\cite{npb592-3,epjc28-515,prd76-074018}
\begin{eqnarray}
\Phi_B=\frac{i}{\sqrt{2N_c}} (p{\hspace{-1.8truemm}/}_B+m_B)\gamma_5\phi_B (k_B),
\label{bmeson}
\end{eqnarray}
where the distribution amplitude $\phi_B$ is of the form
\begin{eqnarray}
\phi_B(x_B,b_B)= N_B x_B^2(1-x_B)^2
\mathrm{exp}\left[-\frac{(x_Bm_B)^2}{2\omega_{B}^2} -\frac{1}{2} (\omega_{B}b_B)^2\right],
\label{phib}
\end{eqnarray}
with $N_B$ the normalization factor. The shape parameters $\omega_B = 0.40 \pm 0.04$ GeV for $B^0$ and $B^\pm$,  
$\omega_{B_s}=0.50 \pm 0.05$ for $B^0_s$, respectively.

The light-cone wave functions for pion and kaon are written as~\cite{jhep9809-005,jhep9901-010,prd71-014015,jhep0605-004}
\begin{eqnarray}
\Phi_{h}=\frac{i}{\sqrt{2N_c}}\gamma_5\left[p{\hspace{-1.8truemm}/}_3\phi^A(x_3)+m^h_0\phi^P(x_3)+
m^h_0(n\hspace{-2.0truemm}/ v\hspace{-1.8truemm}/-1)\phi^T(x_3)\right].
\end{eqnarray}
The distribution amplitudes of $\phi^A(x_3), \phi^P(x_3)$ and $\phi^T(x_3)$ are
\begin{eqnarray}
\phi^A(x_3)&=&\frac{f_{h}}{2\sqrt{2N_c}}6x_3(1-x_3)\left[1+a_1^{h}C_1^{3/2}(t)+a_2^{h}C_2^{3/2}(t)+a_4^{h}C_4^{3/2}(t)\right], \\
\phi^P(x_3)&=&\frac{f_{h}}{2\sqrt{2N_c}}\left[1+(30\eta_3-\frac{5}{2}\rho^2_{h})C_2^{1/2}(t)
-3\big[\eta_3\omega_3+\frac{9}{20}\rho^2_{h}(1+6a_2^{h})\big]C_4^{1/2}(t)\right], \\
\phi^T(x_3)&=&\frac{f_{h}}{2\sqrt{2N_c}}(-t)\left[1+6\left(5\eta_3-\frac{1}{2}\eta_3\omega_3-\frac{7}{20}\rho^2_{h}
-\frac{3}{5}\rho^2_{h}a_2^{h}\right)(1-10x_3+10x_3^2)\right],\quad\;
\end{eqnarray}
with $t=2x_3-1$, $C^{1/2}_{2,4}(t)$ and $C^{3/2}_{1,2,4}(t)$ are Gegenbauer polynomials. The chiral masses $m^h_0$ for pion and 
kaon are $m_0^{\pi}=(1.4 \pm 0.1)$ GeV and $m_0^{K}=(1.6 \pm 0.1)$ GeV as they in Ref.~\cite{prd86-114025}.
The Gegenbauer moments $a_1^{\pi}=0, a_1^K=0.06, a_2^{h}=0.25, a_4^{h}=-0.015$ and the parameters 
$\rho_{h}=m_{h}/m_0^{h}, \eta_3=0.015, \omega_3=-3$ are adopted in the numerical calculation.

For the the differential branching fraction, we have~\cite{PDG-2018}  
\begin{eqnarray}
\frac{d\mathcal{B}}{d\zeta}=\tau_B \frac{q_h q }
 {64 \pi^3 m_B }\overline{|{\cal A}|^2},\label{eqn-diff-bra}
\end{eqnarray}
The magnitude momentum for the bachelor $h$ is
\begin{eqnarray}
q_h=\frac{1}{2}\sqrt{\big[\left(m^2_{B}-m_{h}^2\right)^2 -2\left(m^2_{B}+m_{h}^2\right)s+s^2\big]/s}
\label{def-qh}
\end{eqnarray}
in the center-of-mass frame of the $K^*_0$, where $m_h$ is the mass of the bachelor state.
The direct $CP$ asymmetry ${\mathcal A}_{CP}$ is defined as
\begin{eqnarray}
{\mathcal A}_{CP}=\frac{{\mathcal B}(\bar B\to \bar f)-{\mathcal B}(B\to f)}{{\mathcal B}(\bar B\to \bar f)+{\mathcal B}(B\to f)}
\end{eqnarray}
For errors of the $\mathcal{B}$ and ${\mathcal A}_{CP}$ induced by the parameter ${\mathcal P}\pm\Delta{\mathcal P}$ 
in this work, we employ the formulas
\begin{eqnarray}
\Delta\mathcal{B}=\left|\frac{\partial \mathcal{B}}{\partial{\mathcal P}}\right|\Delta{\mathcal P},\qquad
\Delta{\mathcal A}_{CP}=\left|\frac{\partial \mathcal{A}_{CP}}{\partial{\mathcal P}}\right|\Delta{\mathcal P}=
                          \frac{2(\mathcal{B}\Delta\overline{\mathcal{B}}-\overline{\mathcal{B}}\Delta\mathcal{B})}
                         {(\overline{\mathcal{B}}+\mathcal{B})^2}.
\end{eqnarray}

With the subprocesses $K^{*+}_0\to \{K^0\pi^+, \sqrt{2}K^+\pi^0\}$, $K^{*0}_0\to \{K^+\pi^-, \sqrt{2}K^0\pi^0\}$,
$K^{*-}_0\to \{\bar K^0\pi^-, \sqrt{2}K^-\pi^0\}$ and $\bar K^{*0}_0\to \{K^-\pi^+, \sqrt{2}\bar K^0\pi^0\}$, 
and the $K^{*}_0$ is $K^{*}_0(1430)$ or $K^{*}_0(1950)$, the concerned quasi-two-body decay amplitudes are given as follows:
\begin{eqnarray}
{\cal A}\left(B^+\to K^{*0}_0\pi^+\right)
&=&\frac{G_F}{\sqrt{2}}\big\{V_{ub}^*V_{us}[a_1F^{LL}_{Ah}+C_1M^{LL}_{Ah}]-V_{tb}^*V_{ts}[(a_{4}-\frac{a_{10}}{2})F^{LL}_{Th}
       +(a_6- \frac{a_{8}}{2})F^{SP}_{Th} \nonumber\\
&+&(C_3-\frac{C_9}{2})M^{LL}_{Th}+(C_5-\frac{C_7}{2})M^{LR}_{Th}+(a_4+a_{10})F^{LL}_{Ah}+(a_6+a_8)F^{SP}_{Ah}
\nonumber\\
&+&(C_3+C_9)M^{LL}_{Ah}+(C_5+C_7)M^{LR}_{Ah}]\big\},
\label{eq-das-B+K0pi+}
\\ 
{\cal A}\left(B^+ \to K^{*+}_0\pi^0\right)
&=& \frac{G_F} {2}\big\{V_{ub}^*V_{us}[a_2F^{LL}_{TK^*_0}+C_2M^{LL}_{TK^*_0}+a_1(F^{LL}_{Th}+F^{LL}_{Ah})\nonumber\\
&+&C_1(M^{LL}_{Th}+M^{LL}_{Ah})]-V_{tb}^*V_{ts}[(\frac{3}{2}(a_9-a_7)F^{LL}_{TK^*_0}+\frac{3C_{10}}{2}M^{LL}_{TK^*_0}\nonumber\\
&+&\frac{3C_8}{2}M^{SP}_{TK^*_0}+(a_4+a_{10})(F^{LL}_{Th}+F^{LL}_{Ah})+(a_6+a_8)(F^{SP}_{Th}+F^{SP}_{Ah})\nonumber\\
&+&(C_3+C_9)(M^{LL}_{Th}+M^{LL}_{Ah})+(C_5+C_7)(M^{LR}_{Th}+M^{LR}_{Ah})]\big\},\label{amp11}
\end{eqnarray}
\begin{eqnarray}
{\cal A}\left(B^+ \to K^{*+}_0\bar{K}^0\right)
&=&\frac{G_F} {\sqrt{2}}\big\{V_{ub}^*V_{ud}[a_1F^{LL}_{AK^*_0}+C_1M^{LL}_{AK^*_0}]
       -V_{tb}^*V_{td}[(a_4-\frac{a_{10}}{2})F^{LL}_{TK^*_0}\nonumber\\
&+&(a_6-\frac{a_8}{2})F^{SP}_{TK^*_0}+(C_3-\frac{C_9}{2})M^{LL}_{TK^*_0}+(C_5-\frac{C_7}{2})M^{LR}_{TK^*_0}\nonumber\\
&+&(a_4+ a_{10})F^{LL}_{AK^*_0}+(a_6+a_8)F^{SP}_{AK^*_0}+(C_3+C_9)M^{LL}_{AK^*_0}\nonumber\\
&+&(C_5+C_7)M^{LR}_{AK^*_0}]\big\},\label{amp2}
\\ 
{\cal A}\left(B^+ \to \bar{K}^{*0}_0K^+\right)
&=&\frac{G_F} {\sqrt{2}}\big\{V_{ub}^*V_{ud}[a_1F^{LL}_{Ah}+C_1M^{LL}_{Ah}]
        -V_{tb}^*V_{td}[(a_4-\frac{a_{10}}{2})F^{LL}_{Th}\nonumber\\
&+&(a_6-\frac{a_8}{2})F^{SP}_{Th}+(C_3-\frac{C_9}{2})M^{LL}_{Th}+(C_5-\frac{C_7}{2})M^{LR}_{Th}\nonumber\\
&+&(a_4+ a_{10})F^{LL}_{Ah}+(a_6+a_8)F^{SP}_{Ah}+(C_3+C_9)M^{LL}_{Ah}\nonumber\\
&+&(C_5+C_7)M^{LR}_{Ah}]\big\},\label{amp1}
\\ 
{\cal A}\left(B^0 \to K^{*+}_0\pi^-\right)
&=& \frac{G_F} {\sqrt{2}}\big\{V_{ub}^*V_{us}[a_1F^{LL}_{Th}+C_1M^{LL}_{Th}]-V_{tb}^*V_{ts}[(a_4+a_{10})F^{LL}_{Th}\nonumber\\
&+&(a_6+a_8)F^{SP}_{Th}+(C_3+C_9)M^{LL}_{Th}+(C_5+C_7)M^{LR}_{Th}\nonumber\\
&+&(a_4-\frac{a_{10}}{2})F^{LL}_{Ah}+(a_6-\frac{a_8}{2})F^{SP}_{Ah}+(C_3-\frac{C_9}{2})M^{LL}_{Ah}\nonumber\\
&+&(C_5-\frac{C_7}{2})M^{LR}_{Ah}]\big\},\label{amp14}
\\ 
{\cal A}\left(B^0 \to K^{*0}_0\pi^0\right)
&=& \frac{G_F} {2}\big\{V_{ub}^*V_{us}[a_2F^{LL}_{TK^*_0}+C_2M^{LL}_{TK^*_0}]
        -V_{tb}^*V_{ts}[(\frac{3}{2}(a_9-a_7)F^{LL}_{TK^*_0}\nonumber\\
&+&\frac{3C_{10}}{2}M^{LL}_{TK^*_0}+\frac{3C_8}{2}M^{SP}_{TK^*_0}-(a_4-\frac{a_{10}}{2})(F^{LL}_{Th}+F^{LL}_{Ah})\nonumber\\
&-&(a_6-\frac{a_{8}}{2})(F^{SP}_{Th}+F^{SP}_{Ah})-(C_3-\frac{C_9}{2})(M^{LL}_{Th}+M^{LL}_{Ah})\nonumber\\
&-&(C_5-\frac{C_7}{2})(M^{LR}_{Th}+M^{LR}_{Ah})]\big\},\label{amp13}
\\ 
{\cal A}\left(B^0 \to K^{*+}_0K^- \right)
&=&\frac{G_F} {\sqrt{2}} \big\{V_{ub}^*V_{ud}[a_2F^{LL}_{AK^*_0}+C_2 M^{LL}_{AK^*_0}]
        -V_{tb}^*V_{td}[(a_3+a_9-a_5-a_7)F^{LL}_{AK^*_0}\nonumber\\
&+&(C_4+C_{10})M^{LL}_{AK^*_0}+(C_6+C_8)M^{SP}_{AK^*_0}+(a_3-\frac{a_9}{2}-a_5+\frac{a_7}{2})F^{LL}_{Ah}\nonumber\\
&+&(C_4-\frac{C_{10}}{2})M^{LL}_{Ah}+(C_6-\frac{C_8}{2})M^{SP}_{Ah}]\big\},\label{amp6}
\\ 
{\cal A}\left(B^0 \to K^{*-}_0K^+ \right)
&=&\frac{G_F} {\sqrt{2}} \big\{V_{ub}^*V_{ud}[a_2F^{LL}_{Ah}+C_2 M^{LL}_{Ah}]-V_{tb}^*V_{td}[(a_3+a_9-a_5-a_7)F^{LL}_{Ah}
         \nonumber\\
&+&(C_4+C_{10})M^{LL}_{Ah}+(C_6+C_8)M^{SP}_{Ah}+(a_3-\frac{a_9}{2}-a_5+\frac{a_7}{2})F^{LL}_{AK^*_0}\nonumber\\
&+&(C_4-\frac{C_{10}}{2})M^{LL}_{AK^*_0}+(C_6-\frac{C_8}{2})M^{SP}_{AK^*_0}]\big\},\label{amp5}
\\ 
{\cal A}\left(B^0 \to K^{*0}_0\bar{K}^0\right)
&=&-\frac{G_F} {\sqrt{2}}\big\{V_{tb}^*V_{td}[(a_4-\frac{a_{10}}{2})F^{LL}_{TK^*_0}
       +(a_6-\frac{a_8}{2})(F^{SP}_{TK^*_0}+F^{SP}_{AK^*_0})\nonumber\\
&+&(C_3-\frac{C_9}{2})M^{LL}_{TK^*_0}+(C_5-\frac{C_7}{2})(M^{LR}_{TK^*_0}+M^{LR}_{AK^*_0})+(\frac{4}{3}(C_3+C_4\nonumber\\
&-&\frac{C_9+C_{10}}{2})-a_5+\frac{a_7}{2})F^{LL}_{AK^*_0}+(C_3+C_4-\frac{C_9+C_{10}}{2})M^{LL}_{AK^*_0}\nonumber\\
&+&(C_6-\frac{C_8}{2})(M^{SP}_{AK^*_0}+M^{SP}_{Ah})+(a_3-\frac{a_9}{2}-a_5+\frac{a_7}{2})F^{LL}_{Ah}\nonumber\\
&+&(C_4-\frac{C_{10}}{2})M^{LL}_{Ah}]\big\},\label{amp4}
\end{eqnarray}
\begin{eqnarray}
{\cal A}\left(B^0 \to \bar{K}^{*0}_0K^0\right)
&=&-\frac{G_F} {\sqrt{2}}\big\{V_{tb}^*V_{td}[(a_4-\frac{a_{10}}{2})F^{LL}_{Th}
       +(a_6-\frac{a_8}{2})(F^{SP}_{Th}+F^{SP}_{Ah})\nonumber\\
&+&(C_3-\frac{C_9}{2})M^{LL}_{Th}+(C_5-\frac{C_7}{2})(M^{LR}_{Th}+M^{LR}_{Ah})+(\frac{4}{3}(C_3+C_4\nonumber\\
&-&\frac{C_9+C_{10}}{2})-a_5+\frac{a_7}{2})F^{LL}_{Ah}+(C_3+C_4-\frac{C_9+C_{10}}{2})M^{LL}_{Ah}\nonumber\\
&+&(C_6-\frac{C_8}{2})(M^{SP}_{Ah}+M^{SP}_{AK^*_0})+(a_3-\frac{a_9}{2}-a_5+\frac{a_7}{2})F^{LL}_{AK^*_0}\nonumber\\
&+&(C_4-\frac{C_{10}}{2})M^{LL}_{AK^*_0}]\big\},\label{amp3}
\\ 
{\cal A}\left(B_s^0 \to K^{*-}_0\pi^+\right)
&=& \frac{G_F} {\sqrt{2}}\big\{V_{ub}^*V_{ud}[a_1F^{LL}_{TK^*_0}+C_1M^{LL}_{TK^*_0}]
        -V_{tb}^*V_{td}[(a_4+a_{10})F^{LL}_{TK^*_0}\nonumber\\
&+&(a_6+a_8)F^{SP}_{TK^*_0}+(C_3+C_9)M^{LL}_{TK^*_0}+(C_5+C_7)M^{LR}_{TK^*_0}\nonumber\\
&+&(a_4-\frac{a_{10}}{2})F^{LL}_{AK^*_0}+(a_6-\frac{a_8}{2})F^{SP}_{AK^*_0}+(C_3-\frac{C_9}{2})M^{LL}_{AK^*_0}\nonumber\\
&+&(C_5-\frac{C_7}{2})M^{LR}_{AK^*_0}]\big\},\label{amp16}
\\ 
{\cal A}\left(B_s^0 \to \bar{K}^{*0}_0\pi^0\right)
&=&\frac{G_F} {2}\big\{V_{ub}^*V_{ud}[a_2F^{LL}_{TK^*_0}+C_2M^{LL}_{TK^*_0}]-V_{tb}^*V_{td}[(-a_4-\frac{3a_7}{2}\nonumber\\
&+&\frac{5C_9}{3}+C_{10})F^{LL}_{TK^*_0}-(a_6-\frac{a_8}{2})F^{SP}_{TK^*_0}+(-C_3+\frac{3a_{10}}{2})M^{LL}_{TK^*_0}\nonumber\\
&-&(C_5-\frac{C_7}{2})M^{LR}_{TK^*_0}+\frac{3C_8}{2}M^{SP}_{TK^*_0}-(a_4-\frac{a_{10}}{2})F^{LL}_{AK^*_0} -(a_6\nonumber\\
&-&\frac{a_8}{2})F^{SP}_{AK^*_0}-(C_3-\frac{C_9}{2})M^{LL}_{AK^*_0}-(C_5-\frac{C_7}{2})M^{LR}_{AK^*_0}]\big\},\label{amp15}
\\ 
{\cal A}\left(B_s^0 \to K^{*+}_0K^-\right)
&=&\frac{G_F} {\sqrt{2}}\big\{V_{ub}^*V_{us}[a_1F^{LL}_{Th}+C_1M^{LL}_{Th}+a_2F^{LL}_{AK^*_0}+C_2M^{LL}_{AK^*_0}]\nonumber\\
&-&V_{tb}^*V_{ts}[(a_4+a_{10})F^{LL}_{Th}+(a_6+a_8)F^{SP}_{Th}+(C_3+C_9)M^{LL}_{Th}\nonumber\\
&+&(C_5+C_7)M^{LR}_{Th}+(\frac{4}{3}(C_3+C_4-\frac{C_9+C_{10}}{2})-a_5+\frac{a_7}{2})F^{LL}_{Ah}\nonumber\\
&+&(a_6-\frac{a_8}{2})F^{SP}_{Ah}+(C_3+C_4-\frac{C_9+C_{10}}{2})M^{LL}_{Ah}+(C_5-\frac{C_7}{2})M^{LR}_{Ah}\nonumber\\
&+&(C_6-\frac{C_8}{2})M^{SP}_{Ah}+(a_3+a_9-a_5-a_7)F^{LL}_{AK^*_0}+(C_4+C_{10})M^{LL}_{AK^*_0}\nonumber\\
&+&(C_6+C_8)M^{SP}_{AK^*_0}]\big\},\label{amp10}
\\ 
{\cal A}\left(B_s^0 \to K^{*-}_0K^+\right)
&=&\frac{G_F} {\sqrt{2}}\big\{V_{ub}^*V_{us}[a_1F^{LL}_{TK^*_0}+C_1M^{LL}_{TK^*_0}+a_2F^{LL}_{Ah}+C_2M^{LL}_{Ah}]\nonumber\\
&-&V_{tb}^*V_{ts}[(a_4+a_{10})F^{LL}_{TK^*_0}+(a_6+a_8)F^{SP}_{TK^*_0}+(C_3+C_9)M^{LL}_{TK^*_0}\nonumber\\
&+&(C_5+C_7)M^{LR}_{TK^*_0}+(\frac{4}{3}(C_3+C_4-\frac{C_9+C_{10}}{2})-a_5+\frac{a_7}{2})F^{LL}_{AK^*_0}\nonumber\\
&+&(a_6-\frac{a_8}{2})F^{SP}_{AK^*_0}+(C_3+C_4-\frac{C_9+C_{10}}{2})M^{LL}_{AK^*_0}
      +(C_5-\frac{C_7}{2})M^{LR}_{AK^*_0}\nonumber\\
&+&(C_6-\frac{C_8}{2})M^{SP}_{AK^*_0}+(a_3+a_9-a_5-a_7)F^{LL}_{Ah}+(C_4+C_{10})M^{LL}_{Ah}\nonumber\\
&+&(C_6+C_8)M^{SP}_{Ah}]\big\},\label{amp9}
\end{eqnarray}
\begin{eqnarray}
{\cal A}\left(B_s^0 \to K^{*0}_0\bar{K}^0\right)
&=&-\frac{G_F}{\sqrt{2}}\big\{V_{tb}^*V_{ts}[(a_4-\frac{a_{10}}{2})F^{LL}_{Th}+(a_6-\frac{a_8}{2})(F^{SP}_{Th}+F^{SP}_{Ah})\nonumber\\
&+&(C_3-\frac{C_9}{2})M^{LL}_{Th}+(C_5-\frac{C_7}{2})(M^{LR}_{Th}+M^{LR}_{Ah})+(\frac{4}{3}(C_3+C_4\nonumber\\
&-&\frac{C_9+C_{10}}{2})-a_5+\frac{a_7}{2})F^{LL}_{Ah}+(C_3+C_4-\frac{C_9+C_{10}}{2})M^{LL}_{Ah}\nonumber\\
&+&(C_6-\frac{C_8}{2})(M^{SP}_{Ah}+M^{SP}_{AK^*_0})+(a_3-\frac{a_9}{2}-a_5+\frac{a_7}{2})F^{LL}_{AK^*_0}\nonumber\\
&+&(C_4-\frac{C_{10}}{2})M^{LL}_{AK^*_0}]\big\},\label{amp8}
\\ 
{\cal A}\left(B_s^0 \to \bar{K}^{*0}_0K^0\right)
&=&-\frac{G_F} {\sqrt{2}}\big\{V_{tb}^*V_{ts}[(a_4-\frac{a_{10}}{2})F^{LL}_{TK^*_0}
       +(a_6-\frac{a_8}{2})(F^{SP}_{TK^*_0}+F^{SP}_{AK^*_0})\nonumber\\
&+&(C_3-\frac{C_9}{2})M^{LL}_{TK^*_0}+(C_5-\frac{C_7}{2})(M^{LR}_{TK^*_0}+M^{LR}_{AK^*_0})+(\frac{4}{3}(C_3+C_4\nonumber\\
&-&\frac{C_9+C_{10}}{2})-a_5+\frac{a_7}{2})F^{LL}_{AK^*_0}+(C_3+C_4-\frac{C_9+C_{10}}{2})M^{LL}_{AK^*_0}\nonumber\\
&+&(C_6-\frac{C_8}{2})(M^{SP}_{AK^*_0}+M^{SP}_{Ah})+(a_3-\frac{a_9}{2}-a_5+\frac{a_7}{2})F^{LL}_{Ah}\nonumber\\
&+&(C_4-\frac{C_{10}}{2})M^{LL}_{Ah}]\big\},
\label{eq-das-BsK0K0}
\end{eqnarray}
in which $G_F$ is the Fermi coupling constant, $V$'s are the CKM matrix elements. 
The combinations $a_i$ of the Wilson coefficients are defined as
\begin{eqnarray}
& a_1=C_2+\frac{C_1}{3},\;a_2= C_1+\frac{C_2}{3},\;a_3= C_3+\frac{C_4}{3},\; a_4=C_4+\frac{C_3}{3},\;\;\,
a_5\,= C_5+\frac{C_6}{3},\; \\
& a_6= C_6+\frac{C_5}{3},\;a_7=C_7+\frac{C_8}{3},\;a_8= C_8+\frac{C_7}{3},\;a_9= C_9+\frac{C_{10}}{3},\;
a_{10}= C_{10}+\frac{C_{9}}{3}.
\end{eqnarray}
It should be understood that the Wilson coefficients $C$ and the amplitudes $F$ and $M$ for the factorizable and nonfactorizable 
contributions, respectively, appear in convolutions in momentum fractions and impact parameters $b$.

The general amplitudes for the decays $B\to K^{*}_0 h\to K\pi h$ in the decay amplitudes 
Eq.~(\ref{eq-das-B+K0pi+})-Eq.~(\ref{eq-das-BsK0K0}) are given according to the Fig.~\ref{fig-feyndiag}, the typical Feynman 
diagrams in the PQCD approach. In the following expressions, we will employ $LL$ and $LR$ to denote the contributions from 
$(V-A)(V-A)$ and $(V-A)(V+A)$ operators, respectively. For the contribution from $(S-P)(S+P)$ operators which come from the 
Fierz transformation of the $(V-A)(V+A)$ operators, we will use $SP$ to denote it.
The emission diagrams are depicted in Fig.~\ref{fig-feyndiag} (a) and (c) with $B\to K^*_0$ and $B\to h$ transitions, 
and described as the subscripts $TK^*_0$ and $Th$ in their amplitudes, respectively.  The factorizable and nonfactorizable 
diagrams have been merged in Fig.~\ref{fig-feyndiag}, which could be distinguished easily from the attachments 
of the hard gluons. Those diagrams with two attachments of the hard gluon passed the weak vertex are nonfactorizable diagrams, 
we name their expressions with $M$, while the others are factorizable, and we name their expressions with $F$. 
There are two similar merged annihilation diagrams, the Fig.~\ref{fig-feyndiag} (b) and (d), with the subscripts $AK^*_0$ and 
$Ah$ in their amplitudes, respectively, which demonstrate the $W$ annihilation and $W$-exchange, space-like penguin and time-like 
penguin annihilation-type diagrams.

With the ratio $r_0=m^h_0/m_B$, the amplitudes from Fig.~\ref{fig-feyndiag} (a) are written as
\begin{eqnarray}
F_{TK_0^*}^{LL}   
  &=& 8\pi C_F m^4_B f_{K(\pi)} (\zeta-1)\int dx_B dz\int b_B db_B b db \phi_B(x_B,b_B)\nonumber\\
  &\times&\bigg\{\big[\sqrt{\zeta}(2z-1)(\phi^s+\phi^t)-(z+1)\phi \big]E_{a12}(t_{a1})h_{a1}(x_B,z,b_B,b)\nonumber\\
  &+&\left(\zeta \phi-2\sqrt{\zeta} \phi^s \right) E_{a12}(t_{a2})h_{a2}(x_B,z,b_B,b)\bigg\}, \\
F_{TK_0^*}^{LR}   
  &=&-F_{TK_0^*}^{LL}, \\
F_{TK_0^*}^{SP}   
  &=& 16\pi C_F m^4_B r_0 f_{K(\pi)} \int dx_B dz\int b_B db_B b db \phi_B(x_B,b_B)\nonumber\\
  &\times&\bigg\{\big[\phi[\zeta(2z-1)-1]+\sqrt{\zeta}[z\phi^t-(z+2)\phi^s] \big]E_{a12}(t_{a1})h_{a1}(x_B,z,b_B,b)\nonumber\\
  &+&\left[\phi(2 \zeta-x_B)-2\sqrt{\zeta}\phi^s(\zeta-x_B+1)\right] E_{a12}(t_{a2})h_{a2}(x_B,z,b_B,b) \bigg\},
\\
M_{TK_0^*}^{LL}  
  &=& 32\pi C_F m^4_B/\sqrt{2N_c}(\zeta-1) \int dx_B dz dx_3\int b_B db_B b_3 db_3\phi_B(x_B,b_B)\phi^A \nonumber\\
  &\times& \bigg\{ \big[[\zeta (1-x_3-z)+x_B+x_3-1]\phi+\sqrt{\zeta} z(\phi^s-\phi^t)\big]E_{a34}(t_{a3})
                 h_{a3}(x_B,z,x_3,b_B,b_3)\nonumber\\
  &+&\big[[x_3(1-\zeta)-x_B] \phi+z[\phi-\sqrt{\zeta}(\phi^s+\phi^t)]\big]E_{a34}(t_{a4})h_{a4}(x_B,z,x_3,b_B,b_3) \bigg\}, \\
M_{TK_0^*}^{LR}  
  &=& 32\pi C_F m^4_B r_0/\sqrt{2N_c} \int dx_B dz dx_3\int b_B db_B b_3 db_3\phi_B(x_B,b_B) \nonumber\\
  &\times& \bigg\{ \big[[\zeta(1-x_3)+x_B+x_3-1][\phi+\sqrt{\zeta}(\phi^s-\phi^t)](\phi^P+\phi^T)
                +\sqrt{\zeta}z(\sqrt{\zeta}\phi+\phi^s+\phi^t)\nonumber\\
  &\times&(\phi^T-\phi^P)\big]E_{a34}(t_{a3})h_{a3}(x_B,z,x_3,b_B,b_3)+\big[[(1-\zeta)x_3-x_B][\phi+\sqrt{\zeta}(\phi^s-\phi^t)]
               \nonumber\\
  &\times&(\phi^P-\phi^T)+\sqrt{\zeta}z(\sqrt{\zeta}\phi+\phi^s+\phi^t)(\phi^P+\phi^T)\big]
               E_{a34}(t_{a4})h_{a4}(x_B,z,x_3,b_B,b_3) \bigg\},\quad \\
M_{TK_0^*}^{SP}  
  &=& 32\pi C_F m^4_B/\sqrt{2N_c}(\zeta-1) \int dx_B dz dx_3\int b_B db_B b_3 db_3\phi_B(x_B,b_B)\phi^A \nonumber\\
  &\times& \bigg\{ \big[ [(x_3-1)\zeta-x_B+z-x_3+1]\phi-\sqrt{\zeta} z(\phi^s+\phi^t)\big]E_{a34}(t_{a3})
                 h_{a3}(x_B,z,x_3,b_B,b_3)\nonumber\\
  &+&\big[[x_B+x_3(\zeta-1)]\phi-z\sqrt{\zeta}(\sqrt{\zeta}\phi-\phi^s+\phi^t)\big]E_{a34}(t_{a4})h_{a4}(x_B,z,x_3,b_B,b_3) \bigg\},
\end{eqnarray}
with the color factor $C_F=4/3$. The amplitudes from Fig.~\ref{fig-feyndiag} (b) are written as     
\begin{eqnarray}
F_{AK_0^*}^{LL}  
  &=& 8\pi C_F m^4_B f_B\int dz dx_3\int b db b_3 db_3\nonumber\\
  &\times&\bigg\{\big[ (1-\zeta)(z-1)\phi\phi^A+2\sqrt{\zeta}r_0[(2-z)\phi^s+z\phi^t]\phi^P\big] E_{b12}(t_{b1})
         h_{b1}(z,x_3,b,b_3)\nonumber\\
  &+&\big[(1-\zeta)[x_3(1-\zeta)+\zeta]\phi\phi^A+2\sqrt{\zeta}r_0\phi^s[\big(\zeta(x_3-1)-x_3\big)(\phi^P+\phi^T)
        -(\phi^P-\phi^T)]\big]\nonumber\\
  &\times& E_{b12}(t_{b2})h_{b2}(z,x_3,b,b_3) \bigg\}, 
\end{eqnarray}
\begin{eqnarray}
F_{AK_0^*}^{LR}  
  &=&-F_{AK_0^*}^{LL}, \\
F_{AK_0^*}^{SP}  
  &=& 16\pi C_F m^4_B f_B\int dz dx_3\int b db b_3 db_3\nonumber\\
  &\times&\bigg\{\big[(\zeta-1)\sqrt{\zeta}(z-1)(\phi^s+\phi^t)\phi^A+2r_0\phi\phi^P[\zeta(z-1)-1]\big]E_{b12}(t_{b1})
          h_{b1}(z,x_3,b,b_3)\nonumber\\
  &+&\big[2\sqrt{\zeta}(1-\zeta)\phi^s\phi^A+x_3r_0\phi(\zeta-1)(\phi^P-\phi^T)-2\zeta r_0 \phi\phi^P\big]E_{b12}(t_{b2})\nonumber\\
   &\times&h_{b2}(z,x_3,b,b_3) \bigg\},
\\
M_{AK_0^*}^{LL}  
 &=& 32\pi C_F m^4_B/\sqrt{2N_c} \int dx_B dz dx_3\int b_B db_B b db\phi_B(x_B,b_B)\nonumber\\
 &\times&\bigg\{\big[[{\zeta}^2(1-z-x_3)+\zeta(x_B+2x_3+z-1)-(x_B+x_3)]\phi\phi^A\nonumber\\
 &+&\sqrt{\zeta}r_0[(x_B+(1-\zeta)(x_3-1))(\phi^s-\phi^t)(\phi^P+\phi^T)+z(\phi^s+\phi^t)(\phi^T-\phi^P)+4\phi^s\phi^P]\big] \nonumber\\
 &\times&E_{b34}(t_{b3})h_{b3}(x_B,z,x_3,b_B,b)+\big[({\zeta}^2-1)(z-1)\phi\phi^A+\sqrt{\zeta}r_0[(\zeta(x_3-1)-x_3+x_B)\nonumber\\
 &\times&(\phi^s+\phi^t)(\phi^P-\phi^T)+(z-1)(\phi^s-\phi^t)(\phi^P+\phi^T)]\big]E_{b34}(t_{b4})h_{b4}(x_B,z,x_3,b_B,b)\bigg\}, \\
M_{AK_0^*}^{LR}  
 &=& 32\pi C_F m^4_B/\sqrt{2N_c} \int dx_B dz dx_3\int b_B db_B b db\phi_B(x_B,b_B)\nonumber\\
 &\times&\bigg\{\big[(\zeta-1)(z+1)\sqrt{\zeta}(\phi^s-\phi^t)\phi^A+r_0\phi[(x_3(1-\zeta)+x_B-2)(\phi^P+\phi^T)
            +\zeta z(\phi^T-\phi^P)\nonumber\\
 &+&2\zeta\phi^T]\big]E_{b34}(t_{b3})h_{b3}(x_B,z,x_3,b_B,b)+\big[(1-z)\sqrt{\zeta}(\zeta-1)(\phi^s-\phi^t)\phi^A
           +r_0\phi[(\zeta x_3\nonumber\\
 &-&(x_3-x_B))(\phi^P+\phi^T)+\zeta z(\phi^P-\phi^T)-2\zeta \phi^P]\big]E_{b34}(t_{b4})h_{b4}(x_B,z,x_3,b_B,b)\bigg\}, \\
M_{AK_0^*}^{SP}  
 &=& 32\pi C_F m^4_B/\sqrt{2N_c} \int dx_B dz dx_3\int b_B db_B b db\phi_B(x_B,b_B)\nonumber\\
 &\times&\bigg\{\big[(z\zeta+z-1)(\zeta-1)\phi\phi^A+\sqrt{\zeta}r_0[(\zeta(1-x_3)+x_B+x_3-1)(\phi^s+\phi^t)(\phi^T-\phi^P)\nonumber\\
 &+&z(\phi^s-\phi^t)(\phi^P+\phi^T)-4\phi^s\phi^P]\big]E_{b34}(t_{b3})h_{b3}(x_B,z,x_3,b_B,b)\nonumber\\
 &+&\big[(1-\zeta)[\zeta(z-2)+x_3(\zeta-1)+x_B]\phi\phi^A+r_0\sqrt{\zeta}[(1-z)(\phi^s+\phi^t)(\phi^P-\phi^T)\nonumber\\
 &+&((1-\zeta)x_3+\zeta-x_B)(\phi^s-\phi^t)(\phi^P+\phi^T)]\big]E_{b34}(t_{b4})h_{b4}(x_B,z,x_3,b_B,b)\bigg\},
\end{eqnarray}
The amplitudes from Fig.~\ref{fig-feyndiag} (c) are     
\begin{eqnarray}
F_{Th}^{LL}  
 &=& 8\pi C_F m^4_B  F_{K\pi}(s) /\mu_s  \int dx_B dx_3\int b_B db_B b_3 db_3 \phi_B(x_B,b_B)\nonumber\\
 &\times &\bigg\{\big[(1-\zeta)((\zeta-1)x_3-1)\phi^A-r_0[\big(2x_3(\zeta-1)+\zeta+1\big)\phi^P
     +(\zeta-1)(2x_3-1)\phi^T]\big] \nonumber\\
 &\times&E_{c12}(t_{c1})h_{c1}(x_B,x_3,b_B,b_3)+\left[\zeta(\zeta-1) x_B\phi^A
      +2r_0(\zeta x_B+\zeta-1)\phi^P\right]E_{c12}(t_{c2})  \nonumber\\
 &\times&h_{c2}(x_B,x_3,b_B,b_3)\bigg\},
\end{eqnarray}
\begin{eqnarray}
F_{Th}^{SP}  
 &=& 16\pi C_F m^4_B \sqrt{\zeta}  F_{K\pi}(s)  \int dx_B dx_3\int b_B db_B b_3 db_3 \phi_B(x_B,b_B)\nonumber\\
 &\times &\bigg\{\big[(\zeta-1)\phi^A+r_0[x_3(\zeta-1)(\phi^P-\phi^T)-2\phi^P]\big]E_{c12}(t_{c1})h_{c1}(x_B,x_3,b_B,b_3)\nonumber\\
 &+&\left[(\zeta-1) x_B\phi^A+2r_0(\zeta+x_B-1)\phi^P\right]E_{c12}(t_{c2})h_{c2}(x_B,x_3,b_B,b_3)\bigg\},
\label{exp-figc-Fsp}
\\
M_{Th}^{LL}  
 &=& 32\pi C_F m^4_B/\sqrt{2N_c} \int dx_B dz dx_3\int b_B db_B b db\phi_B(x_B,b_B)\phi \nonumber\\
 &\times&\bigg\{\big[  ({\zeta}^2-1)(x_B+z-1)\phi^A+r_0[\zeta (x_B+z)(\phi^P+\phi^T)
        +x_3(\zeta-1)(\phi^P-\phi^T)\nonumber\\
 &-&2\zeta \phi^P]\big]E_{c34}(t_{c3})h_{c3}(x_B,z,x_3,b_B,b)+\big[(1-\zeta)x_3[(\zeta-1)\phi^A+r_0(\phi^P+\phi^T)]\nonumber\\
 &-&(x_B-z)[(\zeta-1)\phi^A+\zeta r_0(\phi^P-\phi^T)]\big]E_{c34}(t_{c4})h_{c4}(x_B,z,x_3,b_B,b)\bigg\},\\
M_{Th}^{LR}  
 &=& 32\pi C_F m^4_B \sqrt{\zeta}/\sqrt{2N_c} \int dx_B dz dx_3\int b_B db_B b db\phi_B(x_B,b_B)\nonumber\\
 &\times&\bigg\{\big[ (\zeta-1)(x_B+z-1)(\phi^s+\phi^t) \phi^A+r_0[(\zeta(1-x_3)+x_3)(\phi^s-\phi^t)(\phi^P+\phi^T)\nonumber\\
 &+&(x_B+z-1)(\phi^s+\phi^t)(\phi^T-\phi^P) ]\big] E_{c34}(t_{c3})h_{c3}(x_B,z,x_3,b_B,b)\nonumber\\
 &+&\big[(z-x_B)(\phi^s-\phi^t)[(\zeta-1)\phi^A+r_0(\phi^T-\phi^P)]+(\zeta-1)r_0x_3(\phi^s+\phi^t)\nonumber\\
 &\times&(\phi^P+\phi^T)]\big]E_{c34}(t_{c4})h_{c4}(x_B,z,x_3,b_B,b) \bigg\},
\end{eqnarray}
The amplitudes from Fig.~\ref{fig-feyndiag} (d) are     
\begin{eqnarray}
F_{Ah}^{LL}  
 &=& 8\pi C_F m^4_B f_B \int dz dx_3\int b db b_3 db_3 \nonumber\\
 &\times&\bigg\{ \big[[(\zeta-1)[(\zeta-1)x_3+1]\phi \phi^A-2\sqrt{\zeta}r_0 \phi^s[(\zeta-1)x_3(\phi^P-\phi^T)
      +2\phi^P]\big]E_{d12}(t_{d1})\nonumber\\
 &\times&h_{d1}(z,x_3,b,b_3)+ \big[z[2\sqrt{\zeta}r_0(\phi^s+\phi^t)\phi^P+(1-\zeta)\phi\phi^A]
      -2(\zeta-1)\sqrt{\zeta}r_0(\phi^s-\phi^t)\nonumber\\
 &\times&\phi^P\big]E_{d12}(t_{d2})h_{d2}(z,x_3,b,b_3) \bigg\}, \\
F_{Ah}^{LR}  
 &=& -F_{Ah}^{LL},
\\
F_{Ah}^{SP}  
 &=& 16\pi C_F m^4_B f_B \int dz dx_3\int b db b_3 db_3 \nonumber\\
 &\times&\bigg\{ \big[2(1-\zeta)\sqrt{\zeta}\phi^s\phi^A+r_0 \phi[((\zeta-1)x_3+1)(\phi^P+\phi^T)
        +\zeta(\phi^P-\phi^T)]\big]E_{d12}(t_{d1})\nonumber\\
 &\times& h_{d1}(z,x_3,b,b_3)+\left[(1-\zeta)\sqrt{\zeta}z(\phi^s-\phi^t)\phi^A
         +2r_0(\zeta(z-1)+1)\phi\phi^P\right]E_{d12}(t_{d2})\nonumber\\
 &\times&h_{d2}(z,x_3,b,b_3) \bigg\}, 
\end{eqnarray}
\begin{eqnarray}
M_{Ah}^{LL}  
 &=& 32\pi C_F m^4_B/\sqrt{2N_c} \int dx_B dz dx_3\int b_B db_B b db\phi_B(x_B,b_B) \nonumber\\
 &\times&\bigg\{ \big[[(x_B+z-1){\zeta}^2+\zeta-(x_B+z)]\phi\phi^A+\sqrt{\zeta}r_0[(x_3-\zeta(x_3-1))(\phi^s-\phi^t)(\phi^P\nonumber\\
 &+&\phi^T)+(x_B+z-1)(\phi^s+\phi^t)(\phi^T-\phi^P)-4\phi^s\phi^P]\big]E_{d34}(t_{d3})h_{d3}(x_B,z,x_3,b_B,b)\nonumber\\
 &+&\big[ (1-\zeta)[\zeta(x_3-x_B+z-1)-x_3+1]\phi\phi^A-\sqrt{\zeta}r_0[(x_B-z)(\phi^s-\phi^t)(\phi^P+\phi^T)\nonumber\\
 &+&(1-x_3)(\zeta-1)(\phi^s+\phi^t)(\phi^P-\phi^T)]  \big]E_{d34}(t_{d4})h_{d4}(x_B,z,x_3,b_B,b) \bigg\},
\\
M_{Ah}^{LR}  
 &=& 32\pi C_F m^4_B/\sqrt{2N_c} \int dx_B dz dx_3\int b_B db_B b db \phi_B(x_B,b_B)  \nonumber\\
 &\times&\bigg\{\big[ (\zeta-1)\sqrt{\zeta}(x_B+z-2)(\phi^s+\phi^t)\phi^A +r_0\phi[\zeta(x_B+z-1)(\phi^P+\phi^T)\nonumber\\
 &+&(\zeta x_3-x_3-1)(\phi^P-\phi^T)-2\zeta\phi^P] \big]E_{d34}(t_{d3})h_{d3}(x_B,z,x_3,b_B,b)\nonumber\\
 &+&\big[ \sqrt{\zeta}(\zeta-1)(x_B-z)(\phi^s+\phi^t)\phi^A+r_0\phi[\zeta(x_B-z)(\phi^P+\phi^T)\nonumber\\
 &+&(1-x_3)(\zeta-1)(\phi^P-\phi^T)] \big]E_{d34}(t_{d4})h_{d4}(x_B,z,x_3,b_B,b) \bigg\},
\end{eqnarray}
\begin{eqnarray}
M_{Ah}^{SP}  
 &=& 32\pi C_F m^4_B/\sqrt{2N_c} \int dx_B dz dx_3\int b_B db_B b db\phi_B(x_B,b_B)\nonumber\\
 &\times&\bigg\{ \big[(1-\zeta)[\zeta(x_B+z+x_3-2)-x_3+1]\phi\phi^A +\sqrt{\zeta}r_0[(x_B+z-1)(\phi^s-\phi^t)(\phi^P\nonumber\\
 &+&\phi^T)+(x_3\zeta-\zeta-x_3)(\phi^s+\phi^t)(\phi^P-\phi^T)+4\phi^s\phi^P] \big]E_{d34}(t_{d3})h_{d3}(x_B,z,x_3,b_B,b)\nonumber\\
 &+&\big[ (1-{\zeta}^2)(x_B-z)\phi\phi^A+r_0\sqrt{\zeta}[(1-x_3)(\zeta-1)(\phi^s-\phi^t)(\phi^P+\phi^T)\nonumber\\
 &+&(x_B-z)(\phi^s+\phi^t)(\phi^P-\phi^T)] \big]E_{d34}(t_{d4})h_{d4}(x_B,z,x_3,b_B,b) \bigg\},
\end{eqnarray}

\section{PQCD functions}
In this section, we group the functions which appear in the factorization formulas of this work.

With $\bar\zeta=(1-\zeta)$, $\bar x_3=(1-x_3)$ and $\bar z=(1-z)$, the involved hard scales are chosen as
\begin{eqnarray}
t_{a1}&=&\max\left\{m_B\sqrt{z},\; 1/b_B,\; 1/b\right\},\quad
t_{a2}=\max\left\{m_B\sqrt{|x_B-\zeta|}, \;1/b_B, \;1/b\right\},  \\
t_{a3}&=&\max\left\{m_B\sqrt{x_B z},\;m_B\sqrt{z|\bar\zeta\bar x_3-x_B|},\;1/b_B, \;1/b_3 \right\},
\\
t_{a4}&=&\max\left\{m_B\sqrt{x_B z},\;m_B\sqrt{z|x_B-x_3\bar\zeta|},\;1/b_B, \;1/b_3 \right\}, \\
t_{b1}&=&\max\left\{m_B\sqrt{1-z},\; 1/b,\; 1/b_3\right\},\quad
t_{b2}=\max\left\{m_B\sqrt{\zeta+x_3\bar\zeta}, \;1/b,\; 1/b_3\right\}, 
\\
t_{b3}&=&\max\left\{m_B\sqrt{\bar z(\zeta+x_3\bar\zeta)},\;m_B\sqrt{1-z(\bar x_3\bar\zeta-x_B)}, \;1/b_B,\; 1/b \right\}, \\
t_{b4}&=&\max\left\{m_B\sqrt{\bar z(\zeta+x_3\bar\zeta)},\;m_B\sqrt{\bar z|x_B-\zeta-x_3\bar\zeta|}, \;1/b_B, \;1/b \right\}, 
\\
t_{c1}&=&\max\left\{m_B\sqrt{x_3\bar\zeta},\; 1/b_B, \;1/b_3\right\},\quad
t_{c2}=\max\left\{m_B\sqrt{x_B\bar\zeta}, \;1/b_B, \;1/b_3\right\}, \\
t_{c3}&=&\max\left\{m_B\sqrt{x_Bx_3\bar\zeta},\;m_B\sqrt{|1-x_B-z|[x_3\bar\zeta+\zeta]},\;1/b_B,\; 1/b \right\}, \\
t_{c4}&=&\max\left\{m_B\sqrt{x_Bx_3\bar\zeta},\;m_B\sqrt{|x_B-z|x_3\bar\zeta},\;1/b_B, \;1/b \right\}, 
\\
t_{d1}&=&\max\left\{m_B\sqrt{1-x_3\bar\zeta},\; 1/b, \;1/b_3\right\},\quad
t_{d2}=\max\left\{m_B\sqrt{z\bar\zeta}, \;1/b, \;1/b_3\right\}, \\
t_{d3}&=&\max\left\{m_B\sqrt{\bar x_3z\bar\zeta},\;m_B\sqrt{1-(x_3\bar\zeta+\zeta)(1-x_B-z)},\; 1/b_B,\; 1/b \right\}, \\
t_{d4}&=&\max\left\{m_B\sqrt{\bar x_3z\bar\zeta},\;m_B\sqrt{|x_B-z|\bar x_3\bar\zeta},\;1/b_B, \;1/b \right\}. 
\end{eqnarray}

The hard functions are written as 
\begin{eqnarray}
h_{a1}(x_B,z,b_B,b)  
  &=&K_0(m_B\sqrt{x_B z} b_B) \big[\theta(b_B-b)K_0(m_B\sqrt{z} b_B)I_0(m_B\sqrt{z} b) \nonumber\\
  &+& (b \leftrightarrow b_B) \big]S_t(z),\\
h_{a2}(x_B,z,b_B,b)  
  &=&K_0(m_B\sqrt{x_B z} b)S_t(x_B)\nonumber\\
  & \times &\left\{ 
        \begin{array}{ll}\frac{i\pi}{2}\big[\theta(b-b_B)H_0^{(1)}(m_B\sqrt{\zeta-x_B} b)
               J_0(m_B\sqrt{\zeta-x_B} b_B) \\
               \;\quad\quad\; + (b \leftrightarrow b_B) \big],\quad\quad\quad\; x_B<\zeta,\\
                   \big[\theta(b-b_B)K_0(m_B\sqrt{x_B-\zeta} b)
                I_0(m_B\sqrt{x_B-\zeta} b_B) \\
                \;\quad\quad\;  +(b\leftrightarrow b_B)\big],\quad\quad\quad\; x_B\geq\zeta,\\
         \end{array} \right. \\
h_{a3}(x_B,z,x_3,b_B,b_3)  
  &=&\big[\theta(b_B-b_3)K_0(m_B\sqrt{x_Bz} b_B) I_0(m_B\sqrt{x_Bz} b_3)+(b_B\leftrightarrow b_3)  \big]\nonumber\\
  & \times &\left\{ \begin{array}{ll} \frac{i\pi}{2}H_0^{(1)}(m_B\sqrt{z[\bar\zeta\bar x_3-x_B]} b_3),\quad\quad\;\;
     \bar\zeta\bar x_3>x_B,\\
    K_0(m_B\sqrt{z[ x_B-\bar\zeta\bar x_3]} b_3),\quad\quad\quad\quad \bar\zeta\bar x_3\leq x_B,\end{array} \right. 
\\
h_{a4}(x_B,z,x_3,b_B,b_3)  
  &=&\big[\theta(b_B-b_3)K_0(m_B\sqrt{x_Bz} b_B)
     I_0(m_B\sqrt{x_Bz} b_3)+(b_B\leftrightarrow b_3)  \big]\nonumber\\
 & \times &\left\{ \begin{array}{ll}
     \frac{i\pi}{2}H_0^{(1)}(m_B\sqrt{z[x_3\bar\zeta-x_B]} b_3),~\quad\quad x_3\bar\zeta>x_B,\\
    K_0(m_B\sqrt{z[x_B-x_3\bar\zeta]} b_3),~~~\quad\quad\quad x_3\bar\zeta\leq x_B,\end{array} \right.
\\
h_{b1}(z,x_3,b,b_3)  
 &=&\left(\frac{i\pi}{2}\right)^2H_0^{(1)}(m_B\sqrt{\bar z(\zeta+x_3\bar\zeta)} b_3) S_t(z)\nonumber\\
 &\times& \big[\theta(b-b_3)H_0^{(1)}(m_B\sqrt{1-z} b)J_0(m_B\sqrt{1-z} b_3)+ (b \leftrightarrow b_3) \big], 
\\
h_{b2}(z,x_3,b,b_3)  
 &=&\left(\frac{i\pi}{2}\right)^2H_0^{(1)}(m_B\sqrt{\bar z(\zeta+x_3\bar\zeta)} b)S_t(x_3)\big[\theta(b-b_3)\nonumber\\
 &\times&H_0^{(1)}(m_B\sqrt{\zeta+x_3\bar\zeta}b)
     J_0(m_B\sqrt{\zeta+x_3\bar\zeta} b_3)+ (b \leftrightarrow b_3) \big], 
\end{eqnarray}
\begin{eqnarray}
h_{b3}(x_B,z,x_3,b_B,b)  
 &=&\frac{i\pi}{2}K_0(m_B\sqrt{1-z(\bar x_3\bar\zeta-x_B)} b_B)\big[\theta(b_B-b)\nonumber\\
 &\times&H_0^{(1)}(m_B\sqrt{\bar z(\zeta+x_3\bar\zeta)}b_B)J_0(m_B\sqrt{\bar z(\zeta+x_3\bar\zeta)}b)+(b_B\leftrightarrow b)\big],\;\; \\
h_{b4}(x_B,z,x_3,b_B,b)  
 &=&\frac{i\pi}{2}\big[\theta(b_B-b)H_0^{(1)}(m_B\sqrt{\bar z(\zeta+x_3\bar\zeta)} b_B)\nonumber\\
 &\times& J_0(m_B\sqrt{\bar z(\zeta+x_3\bar\zeta)} b)+ (b_B \leftrightarrow b) \big]\nonumber\\
 &\times&\left\{ \begin{array}{ll}
  \frac{i\pi}{2}H_0^{(1)}(m_B\sqrt{\bar z(\zeta+x_3\bar\zeta-x_B)} b_B),~\quad x_B<\zeta+x_3\bar\zeta,\\
   K_0(m_B\sqrt{\bar z(x_B-\zeta-x_3\bar\zeta)} b_B),~~~\quad\quad x_B\geq \zeta+x_3\bar\zeta,\end{array} \right. \\
h_{c1}(x_B,x_3,b_B,b_3)  
 &=& K_0(m_B\sqrt{x_B x_3\bar\zeta} b_B)\big[\theta(b_B-b_3)K_0(m_B\sqrt{x_3\bar\zeta} b_B)\nonumber\\
 & \times & I_0(m_B\sqrt{x_3\bar\zeta} b_3)+ (b_3 \leftrightarrow b_B) \big] S_t(x_3),  
\end{eqnarray}
\begin{eqnarray}
h_{c2}(x_B,x_3,b_B,b_3)  
 &=&h_{c1}(x_3,x_B,b_3,b_B),  \\
h_{c3}(x_B,z,x_3,b_B,b)  
 &=&\big[\theta(b_B-b)K_0(m_B\sqrt{x_B x_3\bar\zeta} b_B)I_0(m_B\sqrt{x_B x_3\bar\zeta} b)+(b_B\leftrightarrow b)\big]\nonumber\\
 &\times&\left\{ \begin{array}{ll}\frac{i\pi}{2}H_0^{(1)}(m_B\sqrt{(1-x_B-z)[x_3\bar\zeta+\zeta]} b),\;\quad x_B+z<1,\\
  K_0(m_B\sqrt{(x_B+z- 1)[x_3\bar\zeta+\zeta]} b),~~\quad\;\quad x_B+z\geq1,\end{array} \right.  \\
h_{c4}(x_B,z,x_3,b_B,b)  
 &=&\big[\theta(b_B-b)K_0(m_B\sqrt{x_B x_3\bar\zeta} b_B)I_0(m_B\sqrt{x_B x_3\bar\zeta} b)+(b_B\leftrightarrow b)  \big]\nonumber\\
 &\times&\left\{ \begin{array}{ll}\frac{i\pi}{2}H_0^{(1)}(m_B\sqrt{x_3(z-x_B)\bar\zeta} b),~\quad\quad x_B<z,\\
    K_0(m_B\sqrt{x_3(x_B-z)\bar\zeta} b),~~~\quad\quad\quad x_B\geq z,\end{array} \right.
\\ 
h_{d1}(z,x_3,b,b_3)  
 &=&\left(\frac{i\pi}{2}\right)^2H_0^{(1)}(m_B\sqrt{\bar x_3z\bar\zeta}b)S_t(x_3)\big[\theta(b-b_3)\nonumber\\
 & \times &H_0^{(1)}(m_B\sqrt{1-x_3\bar\zeta} b)
   J_0(m_B\sqrt{1-x_3\bar\zeta} b_3)+ (b \leftrightarrow b_3) \big],  \\
h_{d2}(z,x_3,b,b_3)  
 &=&\left(\frac{i\pi}{2}\right)^2H_0^{(1)}(m_B\sqrt{\bar x_3z\bar\zeta} b_3)S_t(z)\nonumber\\
 & \times & \big[\theta(b-b_3)H_0^{(1)}(m_B\sqrt{z\bar\zeta} b)J_0(m_B\sqrt{z\bar\zeta} b_3)+ (b \leftrightarrow b_3) \big],  \\
h_{d3}(x_B,z,x_3,b_B,b)  
 &=&\frac{i\pi}{2}K_0(m_B\sqrt{1-x_3(1-x_B-z)\bar\zeta+(x_B+z-1)\zeta} b_B)\nonumber\\
 & \times & \big[\theta(b_B-b)H_0^{(1)}(m_B\sqrt{\bar x_3z\bar\zeta} b_B) J_0(m_B\sqrt{\bar x_3z\bar\zeta} b)
     +(b_B \leftrightarrow b) \big],  \\
h_{d4}(x_B,z,x_3,b_B,b)  
 &=&\frac{i\pi}{2}\big[\theta(b_B-b)H_0^{(1)}(m_B\sqrt{\bar x_3z\bar\zeta} b_B)J_0(m_B\sqrt{\bar x_3z\bar\zeta} b)
     + (b_B \leftrightarrow b) \big]\nonumber\\
 & \times &\left\{ \begin{array}{ll} \frac{i\pi}{2}H_0^{(1)}(m_B\sqrt{\bar x_3(z-x_B)\bar\zeta} b_B),~\quad\quad x_B<z,\\
     K_0(m_B\sqrt{\bar x_3(x_B-z)\bar\zeta} b_B),~~~\quad\quad\quad x_B\geq z,\end{array} \right.
\end{eqnarray}
where $H_0^{(1)}(\chi)=J_0(\chi)+iY_0(\chi)$. The factor $S_t(\chi)$ with the expression~\cite{prd65-014007}
\begin{eqnarray}\label{eq-def-stx}
S_t(\chi)=\frac{2^{1+2c}\Gamma(3/2+c)}{\sqrt{\pi}\Gamma(1+c)}[\chi(1-\chi)]^c,
\end{eqnarray}
resums the threshold logarithms ln$^2\chi$ appearing in the hard kernels to all orders, and the parameter $c$ 
has its expression as $c=0.04Q^2-0.51Q+1.87$ with $Q^2$ the invariant mass square of the final state $f$ 
in the $B\to f$ transition~\cite{prd91-094024,prd80-074024}.

The evolution factors in the factorization expressions are given by
\begin{eqnarray}
E_{a12}(t)&=&\alpha_s(t)  \exp[-S_B(t)-S_{K_0^*}(t)],\\
E_{a34}(t)&=&\alpha_s(t)  \exp[-S_B(t)-S_{K_0^*}(t)-S_{h}]|_{b=b_B}, \\
E_{b12}(t)&=&\alpha_s(t)  \exp[-S_{K_0^*}-S_{h}(t)],\\ 
E_{b34}(t)&=&\alpha_s(t)  \exp[-S_B(t)-S_{K_0^*}(t)-S_{h}]|_{b_3=b}, \\
E_{c12}(t)&=&\alpha_s(t)  \exp[-S_B(t)-S_{K_0^*}(t)],\\
E_{c34}(t)&=&\alpha_s(t)  \exp[-S_B(t)-S_{K_0^*}(t)-S_{h}]|_{b_3=b_B}, \\
E_{d12}(t)&=&E_{b12}(t),\\
E_{d34}(t)&=&E_{b34}(t), 
\end{eqnarray}
in which the Sudakov exponents are defined as
\begin{eqnarray}
S_B&=&s\left(x_B\frac{m_B}{\sqrt2},b_B\right)+\frac53\int^t_{1/b_B}\frac{d\bar\mu}{\bar\mu}
\gamma_q(\alpha_s(\bar\mu)), \\
S_{K_0^*}&=&s\left(z\frac{m_B}{\sqrt2},b\right)+s\left((1-z)\frac{m_B}{\sqrt2},b\right)+
2\int^t_{1/b}\frac{d\bar\mu}{\bar\mu}
\gamma_q(\alpha_s(\bar\mu)), \\
S_{h}&=&s\left(x_3\frac{m_B}{\sqrt2},b_3\right)+s\left((1-x_3)\frac{m_B}{\sqrt2},b_3\right)+
2\int^t_{1/b_3}\frac{d\bar\mu}{\bar\mu}
\gamma_q(\alpha_s(\bar\mu)),
\end{eqnarray}
with the quark anomalous dimension $\gamma_q=-\alpha_s/\pi$.
The explicit form for the  function $s(Q,b)$ is~\cite{prd76-074018}
\begin{eqnarray}
s(Q,b)&=&\frac{A^{(1)}}{2\beta_{1}}\hat{q}\ln\left(\frac{\hat{q}}{\hat{b}}\right)-\frac{A^{(1)}}{2\beta_{1}}\left(\hat{q}-\hat{b}\right)+
\frac{A^{(2)}}{4\beta_{1}^{2}}\left(\frac{\hat{q}}{\hat{b}}-1\right)-\left[\frac{A^{(2)}}{4\beta_{1}^{2}}-\frac{A^{(1)}}{4\beta_{1}}
\ln\left(\frac{e^{2\gamma_E-1}}{2}\right)\right]\;\quad\;\nonumber \\
&\times&\ln\left(\frac{\hat{q}}{\hat{b}}\right)+\frac{A^{(1)}\beta_{2}}{4\beta_{1}^{3}}\hat{q}
\left[\frac{\ln(2\hat{q})+1}{\hat{q}}-\frac{\ln(2\hat{b})+1}{\hat{b}}\right]
+\frac{A^{(1)}\beta_{2}}{8\beta_{1}^{3}}\left[\ln^{2}(2\hat{q})-\ln^{2}(2\hat{b})\right],
\end{eqnarray} 
with the variables are 
\begin{eqnarray}
\hat q\equiv \mbox{ln}[Q/(\sqrt 2\Lambda)],\quad \hat b\equiv
\mbox{ln}[1/(b\Lambda)], \end{eqnarray} and the coefficients
$A^{(i)}$ and $\beta_i$ are 
\begin{eqnarray}
&&\beta_1=\frac{33-2n_f}{12},\quad \beta_2=\frac{153-19n_f}{24}, \quad A^{(1)}=\frac{4}{3}, \nonumber\\
&&A^{(2)}=\frac{67}{9}-\frac{\pi^2}{3}-\frac{10}{27}n_f+\frac{8}{3}\beta_1\mbox{ln}(\frac{1}{2}e^{\gamma_E}),
\end{eqnarray}
where $n_f$ is the number of the quark flavors and $\gamma_E$ is the Euler constant. 


\end{document}